\documentclass[preprints,article,accept,pdftex,moreauthors]{mdpi} 
\pdfoutput=1

\firstpage{1} 
\pubvolume{1}
\issuenum{1}
\articlenumber{0}
\pubyear{2022}
\copyrightyear{2022}
\datereceived{} 
\dateaccepted{} 
\datepublished{} 
\hreflink{https://doi.org/} 

\usepackage{subfig}
\usepackage[per-mode=symbol,separate-uncertainty=true,multi-part-units=single]{siunitx}
\DeclareSIUnit\year{yr}
\usepackage{color,xspace,makecell}
\usepackage{amsfonts,array}
\usepackage{cleveref}
\graphicspath{{./}}
\usepackage{cancel}

\newcommand{\GFC}{\mbox{GF-1}\xspace}
\newcommand{\GFD}{\mbox{GF-2}\xspace}

\newcommand{\pztin}{\texttt{pztIL}\xspace}
\newcommand{\pztout}{\texttt{pztOOL}\xspace}
\newcommand{\thermalin}{\texttt{thermIL}\xspace}
\newcommand{\thermalout}{\texttt{thermOOL}\xspace}
\newcommand{\lastrp}{\texttt{lasTRP}\xspace}

\newcommand{\dd}[1]{\ensuremath{\mathrm{d}#1}\xspace}
\newcommand{\GG}{\ensuremath{\mathrm{G}}\xspace}
\newcommand{\RR}{\ensuremath{\mathrm{R}}\xspace}
\newcommand{\TT}{\ensuremath{\mathrm{T}}\xspace}
\newcommand{\rt}{\ensuremath{\mathrm{[rt]}}\xspace}
\newcommand{\eps}{\ensuremath{\varepsilon\xspace}}

\newcommand{\epsSCF}{\ensuremath{\varepsilon_\mathrm{SCF}}\xspace}
\newcommand{\dtrt}{\ensuremath{\Delta t^\rt}}
\newcommand{\ddtrt}{\ensuremath{\Delta \dot{t}^\rt}}
\newcommand{\depsSCF}{\ensuremath{\dot\varepsilon_\mathrm{SCF}}}
\newcommand{\nuRG}{\ensuremath{\nu_\RR^\GG}}

\usepackage{array}
\newcolumntype{R}[1]{>{\raggedleft\let\newline\\\arraybackslash\hspace{0pt}}m{#1}}
\newcolumntype{Y}{>{\centering\arraybackslash}X}

\usepackage[acronym,nomain]{glossaries}
\usepackage{glossary-mcols}
\makeglossaries
\glsdisablehyper
\setglossarystyle{mcolindex}
\newacronym{AEI}{AEI}{Albert-Einstein Institute}
\newacronym{SC}{SC}{spacecraft}
\newacronym{KBR}{KBR}{K-band Ranging}
\newacronym{MWI}{MWI}{Microwave Instrument}
\newacronym{LRI}{LRI}{Laser Ranging Interferometer}
\newacronym{LRP}{LRP}{Laser Ranging Processor}
\newacronym{TMA}{TMA}{Triple Mirror Assembly}
\newacronym{QPD}{QPD}{Quadrant Photo-Diode}
\newacronym{DWS}{DWS}{Differential Wavefront Sensing}
\newacronym{GRACE}{GRACE}{Gravity Recovery And Climate Experiment}
\newacronym{GFO}{GRACE-FO}{GRACE Follow-On}
\newacronym{LTC}{LTC}{Light Time Correction}
\newacronym{NGGM}{NGGM}{Next Generation Gravity Mission}
\newacronym{SDS}{SDS}{Science Data System}
\newacronym{RLU}{RLU}{Reference Laser Unit}
\newacronym{NPRO}{NPRO}{non-planar ring oscillator}
\newacronym{OBA}{OBA}{Optical Bench Assembly}
\newacronym{PZT}{PZT}{Piezo-Electric Transducer}
\newacronym{OBC}{OBC}{Onboard Computer}
\newacronym{TRP}{TRP}{Thermal Reference Point}
\newacronym{MTS}{MTS}{modulation transfer spectroscopy}
\newacronym{OGSE}{OGSE}{Optical Ground Support Equipment}
\newacronym{TC}{TC}{Thermal Coupling}
\newacronym{IL}{IL}{in-loop}
\newacronym{OOL}{OOL}{out-of-loop}
\newacronym{IPU}{IPU}{MWI Instrument Processing Unit}
\newacronym{USO}{USO}{Ultra-Stable Oscillator}
\newacronym{ASD}{ASD}{amplitude spectral density}
\newacronym{PDH}{PDH}{Pound-Drever-Hall}
\newacronym{FIR}{FIR}{Finite Impulse Response}
\newacronym{TTL}{TTL}{Tilt-To-Length Coupling}
\newacronym{TM}{TM}{telemetry}
\newacronym{MCM}{MCM}{Mass Change Mission}
\newacronym{ULE}{ULE}{ultra-low expansion}
\newacronym{FSR}{FSR}{free spectral range}
\newacronym{GCRF}{GCRF}{geocentric celestial reference frame}
\newacronym{RLAS}{RLAS}{reference laser}
\newacronym{TPR}{TPR}{transponder photoreceiver}
\newacronym{GPS}{GPS}{global positioning system}
\newacronym{CPR}{CPR}{cycles per revolution}
\newacronym{ENBW}{ENBW}{equivalent noise bandwidth}
\newacronym{OG}{OG}{on ground}

\Title{Scale Factor Determination for the GRACE Follow-On Laser Ranging Interferometer including Thermal Coupling}

\TitleCitation{Scale Factor Determination for the GRACE Follow-On Laser Ranging Interferometer including Thermal Coupling}

\Author{Malte Misfeldt $^{1}$*\orcidA{}, Vitali Müller $^{1}$\orcidB{}, Laura Müller $^{1}$, Henry Wegener $^{1}$ and Gerhard Heinzel $^{1}$\orcidC{}}

\AuthorNames{Malte Misfeldt, Vitali Müller, Laura Müller, Henry Wegener and Gerhard Heinzel}

\AuthorCitation{Misfeldt, M.; Müller, V.; Müller, L., Wegener, H., Heinzel, G.}

\address{%
$^{1}$ \quad Max-Planck-Institut für Gravitationsphysik (Albert-Einstein-Institut) and Institut für Gravitationsphysik, Leibniz Universität Hannover, Callinstraße 38, D-30167 Hannover, Germany}

\corres{Correspondence: malte.misfeldt@aei.mpg.de}

\abstract{The GRACE Follow-On satellites carry the very first inter-spacecraft \gls{LRI}. After more than four years in orbit, the \gls{LRI} outperforms the sensitivity of the conventional \gls{MWI}. However, in the current data processing scheme, the \gls{LRI} product still needs the \gls{MWI} data to determine the unknown absolute laser frequency, representing the ``ruler'' for converting the raw phase measurements into a physical displacement in meters.
In this paper, we derive formulas for precisely performing that conversion from the phase measurement into a range, accounting for a varying carrier frequency. Furthermore, the dominant errors due to knowledge uncertainty of the carrier frequency as well as uncorrected time biases are derived. In the second part, we address the dependency of the \gls{LRI} on the \gls{MWI} in the currently employed cross-calibration scheme and present three different models for the \gls{LRI} laser frequency, two of which are largely independent of the \gls{MWI}.
Furthermore, we analyze the contribution of thermal variations on the scale factor estimates and the LRI-MWI residuals. A linear model called \gls{TC} is derived that significantly reduces the differences between \gls{LRI} and \gls{MWI} to a level where the \gls{MWI} observations limit the comparison.}

\keyword{GRACE Follow-On; Laser Ranging Interferometer; Scale Factor; Tone Error; Thermal Coupling} 

\begin{document}

\glsresetall
\section{Introduction}
\label{intro}
The joint U.S.-German gravity space mission \gls{GFO} continues its successful predecessor mission, the \gls{GRACE}.
The twin satellites were launched on the 22nd of May 2018, and the \gls{LRI} was successfully commissioned in mid-June 2018 \cite{Abich2019}. The \gls{GFO} mission was designed to provide data continuity and thus follows the basic concept and design of the predecessor mission. Its main scientific instrument for inter-satellite distance measurements is the \gls{KBR} (or \glsentrylong{MWI}, \glsunset{MWI}\gls{MWI}) together with an accelerometer on each spacecraft to determine non-gravitational accelerations acting on the two spacecraft for later removal in the data processing. 
Global observations of Earth's gravitational potential and its variations from space allow valuable insights into the hydrological cycle, including rainfall, droughts, ice-melting, and sea-level rise \cite{Landerer2020}.

New to \gls{GFO} is the \gls{LRI}, a technology demonstrator to prove the feasibility of laser interferometry for distance measurements between two spacecraft flying a few hundred kilometers apart. The \gls{LRI} shows drastically increased precision compared to the \gls{KBR} instrument \cite{Abich2019,GhobadiFar2020}. 
Based on the flawless in-orbit operation for over four years and without any degradation observed so far, the \gls{LRI} technology is now being adopted to serve as the primary instrument in future missions like \gls{NGGM}, and GRACE-I(carus)/\gls{MCM} \cite{Conklin2020,Massotti2021,Nicklaus2019,Nicklaus2022}. Evolving from a demonstrator to a primary instrument will include changes concerning reliability and redundancy. Moreover, the success in demonstrating inter-spacecraft laser interferometry was a milestone for the space-based gravitational wave observatory LISA \cite{AmaroSeoane2017}.

The \gls{LRI} has shown very low noise in the inter-satellite ranging measurement of about \SI{0.3}{\nano\meter\per\sqrt{\hertz}} at Fourier frequencies of \SI{1}{\hertz} \cite{Abich2019}, which is about three orders of magnitude below the noise of the \gls{MWI}. However, the conversion factor between the raw phase measurement of the heterodyne interferometer and the desired displacement is needed to form the \gls{LRI} ranging signal. This conversion factor is the wavelength $\lambda = c_0/\nu_\RR$, with $c_0$ denoting the speed of light in vacuum. 
The \gls{LRI} laser frequency on the reference satellite $\nu_\RR$ is actively stabilized to a resonance of an optical reference cavity using the \gls{PDH} technique. 
The variations in the cavity's resonance frequency mainly depend on the resonator's thermal stability, and the frequency's absolute value can not be measured directly in flight. Therefore, the current data processing scheme foresees a cross-calibration of \gls{LRI} and \gls{KBR} to determine the relative scaling between the \gls{KBR} range and \gls{LRI} range, using an initial estimate $\nu_0$ for the laser frequency. Through rescaling the initial value $\nu_0$, the actual laser frequency is approximated as 
\begin{equation}
	\nu_\mathrm{estim.} = \frac{1}{1+\epsSCF}\nu_0\ . \label{eq::scf_definition}
\end{equation}

This paper aims to investigate approaches to decrease the dependency of the \gls{LRI} data on \gls{KBR} data in case the latter is unavailable and to study the performance of a possible \gls{LRI}-only \gls{NGGM}. Therefore, we develop different models for estimating the \gls{LRI} laser frequency in-flight. Since these models do not achieve the same level of residuals as the cross-calibration with \gls{KBR}, a potential influence of thermal variations into the ranging data is investigated and modeled.

In \cref{sec::theory}, we briefly cover the working principle of the \gls{LRI} and introduce the instrument's most important optical and radiofrequency observables. 
The two dominant error sources in the \gls{LRI}-derived range, which occur in the actual data processing, are discussed in \cref{sec::errorCouplingModel}. 
Flight data processing is the topic of \cref{sec::SCFdetermination}, where we analyze the KBR-LRI cross-calibration method and discuss an observed frequency change of the cavity resonance onboard \GFC.
The derivation and calibration of a telemetry-based absolute laser frequency model is presented in \cref{sec::lasertelemetry,sec::ongroundCalibration} and an empirical correction to this model is derived in \cref{sec::empiricalModel}. 
In \cref{sec::inflightapplication}, we use the frequency models to derive three independent LRI1B-equivalent ranging data sets, which are then compared to each other and the KBR. In the end, \cref{sec::toneErrors} focuses on minimizing variations in the relative scale and timeshift of LRI and KBR by reducing thermally-induced measurement errors, which predominantly manifest as tone errors. Coupling factors are derived to model this effect and subtract it from the ranging data. 
Available technologies to determine the absolute laser frequency for future gravity missions are discussed in \cref{sec::discussion}, and the findings are summarized in \cref{sec::conclusion}.

\section{Working Principle of LRI} \label{sec::theory}
\begin{figure}
	\centering
	\includegraphics[width=0.75\linewidth]{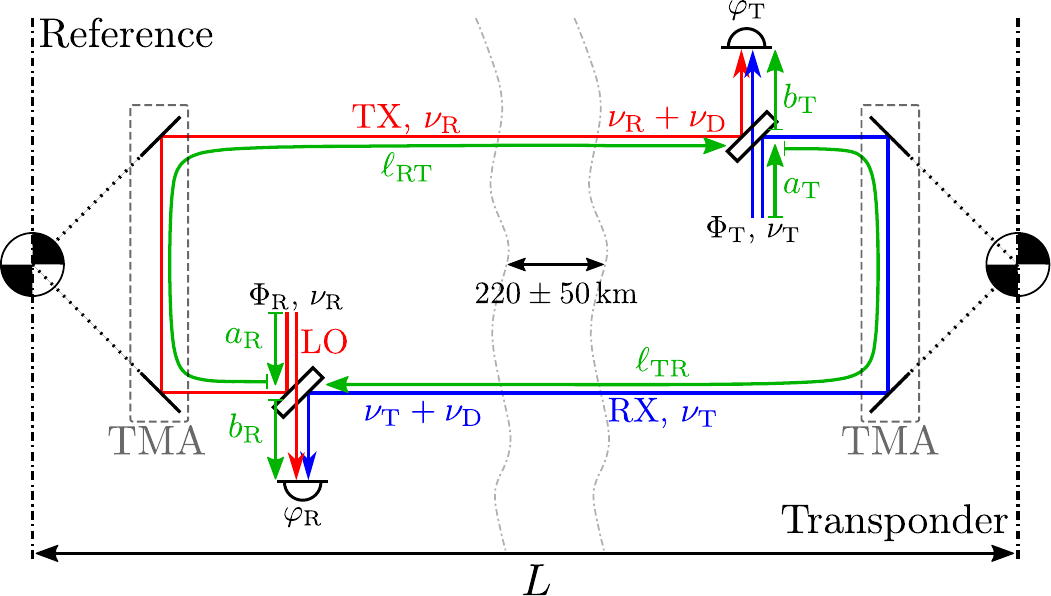}
	\caption{Simplified light paths and frequencies within the LRI. See the main text for an explanation.}
	\label{fig::scheme}
\end{figure}

The \gls{LRI} is set up in an active-transponder configuration \cite{Sheard2012}, the principle of which is shown in \cref{fig::scheme}. Both spacecraft have identical hardware, including a laser, and both receive and emit light. They are equipped with photoreceivers to measure the interference between the incoming and local light fields (shown in red and blue in \cref{fig::scheme}). The \gls{LRI} is a heterodyne Mach-Zehnder type interferometer, meaning that the two interfering light fields have slightly different optical frequencies, which produces an interference beatnote at the difference frequency. This beatnote frequency is roughly \SI{10}{\mega\hertz} for the \gls{LRI}.

On the reference side, the laser frequency $\nu_\RR$ is stabilized utilizing an optical reference cavity using the \gls{PDH} technique \cite{Drever1983}. The residual frequency fluctuations $\delta \nu_\RR$ were required to be below $\SI{30}{\hertz\per\sqrt\hertz}$ for Fourier frequencies above \SI{10}{\milli\hertz}, with a relaxation towards lower Fourier frequencies \cite{Sheard2012,Thompson2011}. The actual in-flight noise, expressed as \gls{ASD}, is well below the requirement and in the order of
\begin{equation}
    \mathrm{ASD}[\delta \nu_\RR](f) \approx  \frac{\num{e-15}\cdot\nu_\RR}{\sqrt{\si{\hertz}}}  \sqrt{\frac{f}{\SI{1}{\hertz}}} \approx \num{0.3}\frac{\si{\hertz}}{\sqrt{\si{\hertz}}} \sqrt{\frac{f}{\SI{1}{\hertz}}} \label{eq::LFN} 
\end{equation}
at frequencies above \SI{200}{\milli\hertz} \cite{Abich2019}, where it is dominant and directly apparent in the measured signal due to the lack of other signals at such high frequencies. Assessing the frequency stability at lower frequencies is difficult due to the dominant ranging signal arising from gravitational and non-gravitational differential forces acting on the satellites (cf. black trace in \cref{fig::1ppm}, page \pageref{fig::1ppm}).

On the reference spacecraft, the laser light is split at the beamsplitter into a local oscillator (LO) part and into the transmit (TX) beam (cf.~\cref{fig::scheme}). The \gls{TMA} routes the TX beam around cold-gas tanks (not shown) towards the distant spacecraft. The emitted frequency is Doppler-shifted when received on the transponder due to the relative motion of the two spacecraft. The relative velocity of the two spacecraft is below \SI{\pm 2.5}{\meter\per\second}, which translates into a one-way Doppler shift of $\nu_\mathrm{D} < \SI{\pm 2.5}{\mega\hertz}$ \cite{Sheard2012}. 

The transponder unit employs a frequency-locked loop with a \SI{10}{\mega\hertz} offset, meaning that the frequency $\nu_\TT$ is controlled with high gain and bandwidth such that the beatnote at the photodetector between the local and received, Doppler-shifted light stays at $f_\mathrm{off}=\SI{10}{\mega\hertz}$. This enforces that the transponder laser frequency $\nu_\TT$ is the sum of received Doppler-shifted reference frequency $\nu_\RR^\prime = \nu_\RR + \nu_\mathrm{D}$ and $f_\mathrm{off}$.
The trans\-pon\-der is to send back amplified laser light to the reference with a well-defined and known optical frequency (and phase). Since the transponder is in \SI{220(50)}{\kilo\meter} distance, it only receives a fraction of the initially emitted light power (in the order of nanowatts). The amplified and offset-locked beam travels back to the reference side, interferes again, and the beatnote between the (once more) Doppler-shifted transponder and local reference frequency $f_\RR = \nu_\TT^\prime  - \nu_\RR = 2\nu_\mathrm{D}+f_\mathrm{off}$ contains the desired ranging information, encoded in the Doppler shift $\nu_\mathrm{D}$.

\section{Error Coupling in the Range Measurement} \label{sec::errorCouplingModel}
The previous section provided a descriptive picture of the \gls{LRI} working principle through the beam's frequencies. However, the \gls{LRI} actually measures the differential phase of the two interfering light beams given by the time integral of the beatnote frequency. To describe the phase observables in a relativistic framework, we now follow the approach of Yan et al. \cite{Yan2021} to assess potential relativistic effects on the scale factor. We favor this description in terms of phase since that is invariant in the context of general relativity, i.\,e., independent of the coordinate system, in contrast to the frequency. 

The conversion of the measured differential phase 
\begin{equation}
	\varphi_\mathrm{LRI} = \varphi_\TT - \varphi_\RR
\end{equation}
to the range observable in a relativistic framework is given in \cite{Yan2021,Mueller2022} and is omitted here for brevity.
The LRI range reads
\begin{align}
	\rho_\mathrm{LRI}^\mathrm{raw}(t) &= \frac{c_0}{2}\int_0^t \frac{\dd{\varphi_\mathrm{LRI}(t^\prime)}}{\dd{t^\prime}}\frac{1}{\nu_\RR^\GG(t^\prime-\Delta t^\rt(t^\prime))} -\left(\frac{\nu_\RR^\GG(t^\prime)}{\nu_\RR^\GG(t^\prime-\Delta t^\rt(t^\prime))}-1\right)\,\dd{t^\prime} \label{eq::rawRange1} \\
	&= \frac{c_0}{2} \left(\Delta t^\rt(t)-\Delta t^\rt(0)\right) + \mathrm{errors} \ . \label{eq::rho_lri_approx}
\end{align}
\Cref{eq::rawRange1} provides a recipe to compute the raw biased range as a function of the coordinate time $t$, which is available after precise orbit determination. The roundtrip propagation time is in the order of $\dtrt\approx 2\cdot\SI{220}{\kilo\meter}/c_0\approx\SI{1.46}{\milli\second}$ with the speed of light $c_0$ and the absolute laser frequency in the coordinate frame is given by $\nu^\GG_\RR$. The relation between the frequency of the laser source $\nu_\RR$ and the apparent frequency in the Earth-centered \gls{GCRF} system $\nu_\RR^\GG$ is 
\begin{equation}
	\nu_\RR^\GG = \nu_\RR\cdot\frac{\dd{\tau_\RR}}{\dd{t}}\ ,
\end{equation}
where $\tau_\RR$ is the proper time of the reference spacecraft and $t$ is the coordinate time in the \gls{GCRF}. The distinction between those is discussed in \cite{Mueller2022} and a neglection yields a tone error in the order of \SI{1}{\micro\meter\,rms} at 1/rev. The first term in the integral resembles the well known relation
\begin{equation}
	\rho(t) = \frac{c_0}{\nu}\varphi(t)
\end{equation}
and the second term accounts for the effect of a varying frequency $\nu_\RR^\GG(t)$.

\Cref{eq::rho_lri_approx} in turn provides the physical meaning of $\rho_\mathrm{LRI}^\mathrm{raw}$ as a time-of-flight measurement, whereby the errors include \glsentrylong{TTL} \cite{Wegener2020}, laser frequency noise \cite{Abich2019} and others.

The representation until here is, however, neglecting some error sources. We now derive a model for two error sources, namely a mismodeling of the laser frequency, which can be expressed through a scale factor, and secondly from clock errors.
\Cref{eq::rawRange1} shows that the inter-satellite biased range can be reconstructed from phase measurements if the conversion factor, given by the absolute frequency $\nu_\RR^\GG(t)$ or wavelength $\lambda(t)=c_0 / \nu_\RR^\GG(t)$, is known. If we consider errors in the knowledge of the frequency, given as the difference between estimated and true frequency $\nu_\mathrm{R,est}^\GG(t) - \nu_\RR^\GG(t)$, this error is typically expressed as a scale factor 
\begin{equation}
	\label{eq::epsSCF}
	\epsSCF(t) = \frac{\nu_\mathrm{R,est}^\GG(t) - \nu_\RR^\GG(t)}{\nu_\RR^\GG(t)} 
\quad\Leftrightarrow\quad
	\nu_\RR^\GG(t) = \frac{\nu_\mathrm{R,est}^\GG(t)}{1+\epsSCF(t)}\ .
\end{equation}
By applying the replacement $\nu_\RR^\GG(t) \rightarrow \nu_{\RR}^\GG(t)/(1+\epsSCF(t))$ to \cref{eq::rawRange1}, we obtain an expression for the estimated range $\rho_\mathrm{LRI}^\mathrm{raw,est}$. In the following, we compute the error of this estimated range. For better readability, we drop the time dependency of terms evaluated at the measurement epoch $t^\prime$ in the integral.
\newcommand{\myint}{\frac{c_0}{2}\int_0^t}
\newcommand{\tprime}{t^\prime}
\newcommand{\tmdt}{\tprime\!-\!\dtrt}
\begin{align}
	&\rho_\mathrm{LRI}^\mathrm{raw,est}(t)-\rho_\mathrm{LRI}^\mathrm{raw}(t) \notag \\
	&\qquad= \myint \frac{\nuRG+\epsSCF(\tmdt)\cdot\dot\varphi_\mathrm{LRI}}{\nuRG(\tmdt)} - \frac{(1+\epsSCF(\tmdt))\nuRG}{(1+\epsSCF)\nuRG(\tmdt)}\ \dd{\tprime} \label{eq::error_coup1}\\
	&\qquad\approx\myint \epsSCF(\tmdt)\,\ddtrt - \epsSCF(\tmdt) + \frac{\nuRG\epsSCF}{\nuRG(\tmdt)} \ \dd{\tprime} \label{eq::error_coup2} \\
	&\qquad\approx\myint \epsSCF\ddtrt + \depsSCF\dtrt - \epsSCF + \frac{\nuRG\epsSCF}{\nuRG(\tmdt)}\ \dd{\tprime} \label{eq::error_coup3}\\
	&\qquad\approx \epsSCF(t)\cdot\frac{c_0}{2}\dtrt(t) + \myint \epsSCF\left(\frac{\dot\nu_\RR^\GG}{\nuRG}\dtrt+1\right) -\epsSCF\ \dd{\tprime} \label{eq::error_coup4} \\
	&\qquad\approx \epsSCF(t)\cdot L(t) \label{eq::delta-epsSCF}
\end{align}
where we used some approximations justified below to obtain the simple result. The approximation in \cref{eq::error_coup2} is based on the relation $1/(1+\epsSCF) \approx 1 - \epsSCF$ together with the definition of the phase derivative in \cite[eq. 27]{Mueller2022} and we dropped a second order term in $(\epsSCF)^2$. \Cref{eq::error_coup3} uses $\epsSCF(t-\Delta t^\rt) \approx \epsSCF(t) -\Delta t^\rt \depsSCF(t)$ and neglects product terms of $\dtrt\cdot\ddtrt$. The result of \cref{eq::error_coup4} employed the same type of approximations, namely $1/\nu_\RR^\GG(\tmdt) \approx (1+\dot{\nu}_\RR^\GG/\nuRG \cdot \dtrt)/\nuRG$. To solve the integral, we omitted product terms of $\dot{\nu}_\RR^\GG/\nu_\RR^\GG$ with $\epsSCF$ or $\depsSCF$, because these describe a second order cross-coupling between scale error and fractional true frequency change that is expected to be negligible. $L(t) = c_0\dtrt(t)/2$ denotes the absolute distance between the spacecraft and the error coupling $\epsSCF(t)\cdot L(t)$ resembles the well-known influence of (fractional) laser frequency variations into the range measurement \cite{Sheard2012}, which can be regarded as a scale factor error. 

The second error contributor that we address is a potential timeshift $\zeta$ of the measured \gls{LRI} data, arising from unmodeled internal delays of the LRI. At startup, the \gls{LRI} time is initialized via the \gls{OBC}, which introduces a delay of \SI{1.5}{\second} at maximum \cite{Level1UserHandbook}, although we only observed values below \SI{1.0}{\second}. To compensate for this delay, the differences of \gls{LRI} time and \gls{IPU} time are measured regularly (called the datation report) and are used to correct the \gls{LRI} time tags. However, a small deviation $\zeta$ may remain, even after this subtraction. We linearize the effect of this potential timeshift to first order as
\begin{equation}
    \rho_\mathrm{LRI}^\mathrm{inst}(t+\zeta) \approx \rho_\mathrm{LRI}^\mathrm{inst}(t) + \zeta\cdot\dot\rho(t) \label{eq::timeshift}\ ,
\end{equation}
where we use the approximate range rate $\dot{\rho} \approx \dot{\rho}_\textrm{LRI}^\mathrm{raw} \approx \dot{\rho}_\mathrm{LRI}^\mathrm{inst}$ for terms that describe a small error coupling and where the highest precision in $\dot\rho$ is not required.

\begin{figure}[tb]
	\centering
	\includegraphics[width=\linewidth]{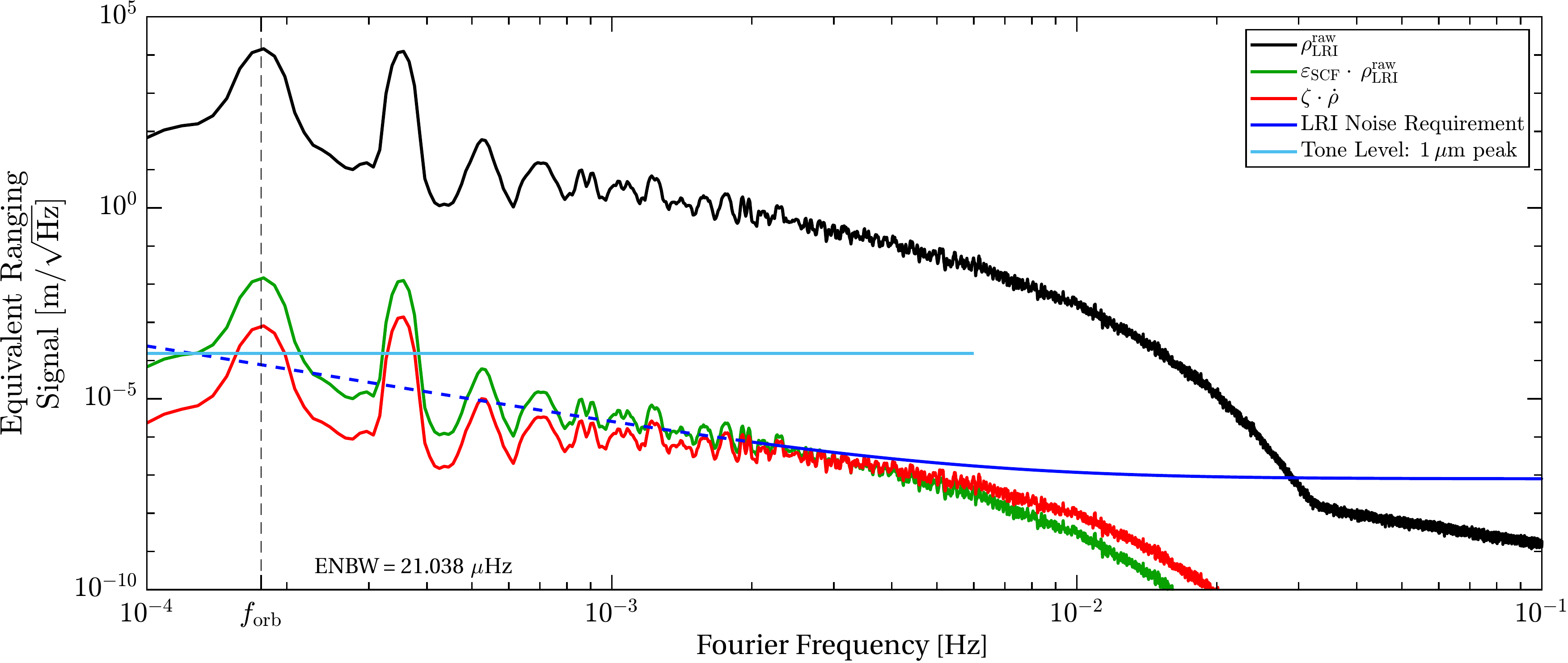}
	\caption{Typical amplitude spectral density of the LRI ranging signal (black) and effective errors arising from a static scale factor error $\epsSCF=10^{-6}$ (green) and timeshift $\zeta = \SI{50}{\micro\second}$ (red, cf. \cref{eq::timeshift}). Also shown is the noise requirement of LRI (blue), which is strictly applicable only for frequencies above \SI{2}{\milli\hertz}, but it was extrapolated towards lower frequencies (blue dashed segment). The light blue line denotes a \SI{1}{\micro\meter} tone amplitude (computed via the \gls{ENBW}). The ranging measurement is dominated by laser frequency noise at highest frequencies (above \SI{30}{\milli\hertz}), and by the differential gravitational and non-gravitational forces below.}
	\label{fig::1ppm}
\end{figure}
\Cref{fig::1ppm} illustrates the significance of the static scale factor error $\epsSCF = 10^{-6}$ (green) and timeshift $\zeta = \SI{50}{\micro\second}$ (red), which represent the common orders of magnitude in current \gls{LRI} data processing. The effects of these exceed or are close to the \gls{LRI} noise requirement for frequencies between $\SIrange[range-units=single,range-phrase=\dots]{0.6}{3}{mHz}$, indicating that the scale factor and timeshift need to be known to better precision, e.\,g., at the level of \numrange{e-7}{e-8} for the scale and at a level of a few microseconds or better for the timeshift. Fourier frequencies below $\SI{0.6}{mHz}$ are dominated by sinusoidal errors at integer multiples of the orbital frequency $f_\mathrm{orb} \approx \SI{0.18}{\milli\hertz}$. These peaks need to be compared to tone error requirements with the unit of a meter (rms or peak) instead of spectral densities with the unit of \si{\meter\per\sqrt{\Hz}}. An indicative level for tone errors is $\SI{1}{\micro\meter}$ (peak), which is a requirement for the \gls{MWI} at 2/rev frequency in GRACE-FO \cite{Kornfeld2019}, is shown in light blue. The displayed errors exceed the one micron tone level by approximately two orders of magnitude. 

By combining the effects due to a scale factor error (\cref{eq::delta-epsSCF}) and an uncompensated timeshift (\cref{eq::timeshift}), we obtain the error of the \gls{LRI} measured range w.r.t. the truth as
\begin{equation} 
	\rho_\mathrm{LRI}^\mathrm{inst}(t) - \rho^\mathrm{inst}_\mathrm{true}(t) \approx  \epsSCF(t) \cdot L(t) + \zeta \cdot \dot{\rho}(t)\ .
	\label{eq::rho_LRI-rho_ref}
\end{equation}

\section{Scale Factor Determination} \label{sec::SCFdetermination}
The scale factor \epsSCF implicitly defined in \cref{eq::scf_definition} provides an estimate $\nu_\mathrm{R,est}$ for the actual laser frequency $\nu_\RR$ of the LRI reference unit, which in turn is needed for accurately converting the phase measurement into a range in meter. Here we present three approaches to either directly calculate the absolute laser frequency $\nu_\RR$ or through the scale factor \epsSCF that is related to absolute laser frequency through \cref{eq::epsSCF}.

Since \gls{GFO} hosts the \gls{KBR} and \gls{LRI}, which are designed to measure the same quantity in parallel, the obvious way to obtain the \gls{LRI} scale $\epsSCF$ (or frequency $\nu_\RR$) is to compare the ranging data of the two instruments. We define the instantaneous \gls{KBR} range as
\begin{equation}
	\rho_\mathrm{KBR}^\mathrm{inst}(t) = \rho_\mathrm{KBR}^\mathrm{raw}(t) + \rho_\mathrm{KBR}^\mathrm{LTC}(t) + \rho_\mathrm{KBR}^\mathrm{AOC}(t)\ .
\end{equation}
The three quantities on the right-hand side are
the ionosphere-free K/Ka-band range $\rho_\mathrm{KBR}^\mathrm{raw}$, the \gls{KBR} \gls{LTC} and the antenna offset correction (AOC), which are regarded as error-free here. They are given in the KBR1B data product \cite{Level1UserHandbook}. Ultimately, daily arcs of \gls{LRI} phase measurements can be calibrated against the \gls{KBR} ranging data, i.\,e., by minimizing the KBR-LRI residuals as
\begin{equation}
	\left\|\rho_\mathrm{KBR}^\mathrm{inst}(t) - \lambda_\mathrm{est}^\mathrm{SDS}\cdot\varphi_\mathrm{LRI}(t+\zeta) - \rho^\mathrm{LTC}_\mathrm{LRI}\right\| \rightarrow 0 \label{eq::eps_LSQ}
\end{equation} 
using a daily constant laser wavelength $\lambda_\mathrm{est}^\mathrm{SDS}$ and timeshift $\zeta$ as fit parameters. The \gls{KBR} scale factor error can be regarded as negligible since the relevant \gls{USO} frequency is determined during precise orbit determination by referencing it to \gls{GPS}. The \gls{USO} fractional frequency varies by about \num{e-11}, mainly at 1/rev \cite[Fig. 1]{Mueller2022}, and we assume the knowledge error to be even smaller. The processing of daily chunks of data essentially decomposes the scale factor \epsSCF into a static and time-variable part as 
\begin{equation}
	\epsSCF(t) = \langle\epsSCF\rangle + \delta\epsSCF(t)\ , \label{eq::epsSCF_decomposed}
\end{equation}
of which only the static part $\langle\epsSCF\rangle$ is determined separately on every day with discontinuities at the day boundary.

This cross-calibration scheme is the official processing strategy employed by the \gls{SDS} for the LRI1B data product in version 04, where a conversion factor from phase to range and a timeshift $\zeta$ is estimated once per day. The scale  $\langle\epsSCF^\mathrm{SDS}\rangle$ is reported in the ionospheric correction (\texttt{iono\_corr}) column of the LRI1B files, whereas the timeshift $\zeta$ is applied through LLK1B.
The scale value relates to the laser frequency and wavelength estimates of the reference laser through
\begin{align} 
	&\lambda_\mathrm{est}^\mathrm{SDS}(t_\mathrm{daily}) = (1+\langle\epsSCF^\mathrm{SDS}\rangle(t_\mathrm{daily})) \cdot \lambda_0
\intertext{and}
	&\nu_\mathrm{est}^\mathrm{SDS}(t_\mathrm{daily}) = \frac{\nu_0}{\left(1+\langle\epsSCF^\mathrm{SDS}\rangle(t_\mathrm{daily})\right)}\ , \label{eq::nuEps}
\end{align}
where $\nu_0$ is a nominal frequency for the LRI lasers, given in the documentation as $\nu_0 = \SI{281616393}{\mega\hertz}$ and $\nu_0 = \SI{281615684}{\mega\hertz}$ for GF-1 and GF-2, respectively \cite{Level1UserHandbook}. The two nominal values were determined pre-flight, but do not represent the best knowledge of the actual frequency. The corresponding nominal wavelength is $\lambda_0 = c_0 / \nu_0$. \Cref{eq::nuEps} is in principle a reformulation of \cref{eq::epsSCF,eq::epsSCF_decomposed}, however, the time-varying part $\delta\epsSCF(t)$ is neglected by the processing of daily segments. 
Therefore, \cref{eq::rawRange1} simplifies to
\begin{equation}
	\rho_\mathrm{LRI}^\mathrm{SDS}(t) =  \frac{c_0}{2}\frac{(1+\langle\epsSCF^\mathrm{SDS}\rangle)}{\nu_0} \cdot\varphi_\mathrm{LRI}(t)\ . \label{eq::SDS_eps2}
\end{equation}

\def\cavityDriftDecayRate{\SI{4.006e-8}{\per\second}\xspace}
\def\cavityDriftInitial{\num{-1.190e-7}\xspace}
\def\cavityDriftAsymptote{\num{2.221e-6}\xspace}
\begin{figure}
	\centering
	\includegraphics[width=0.99\linewidth]{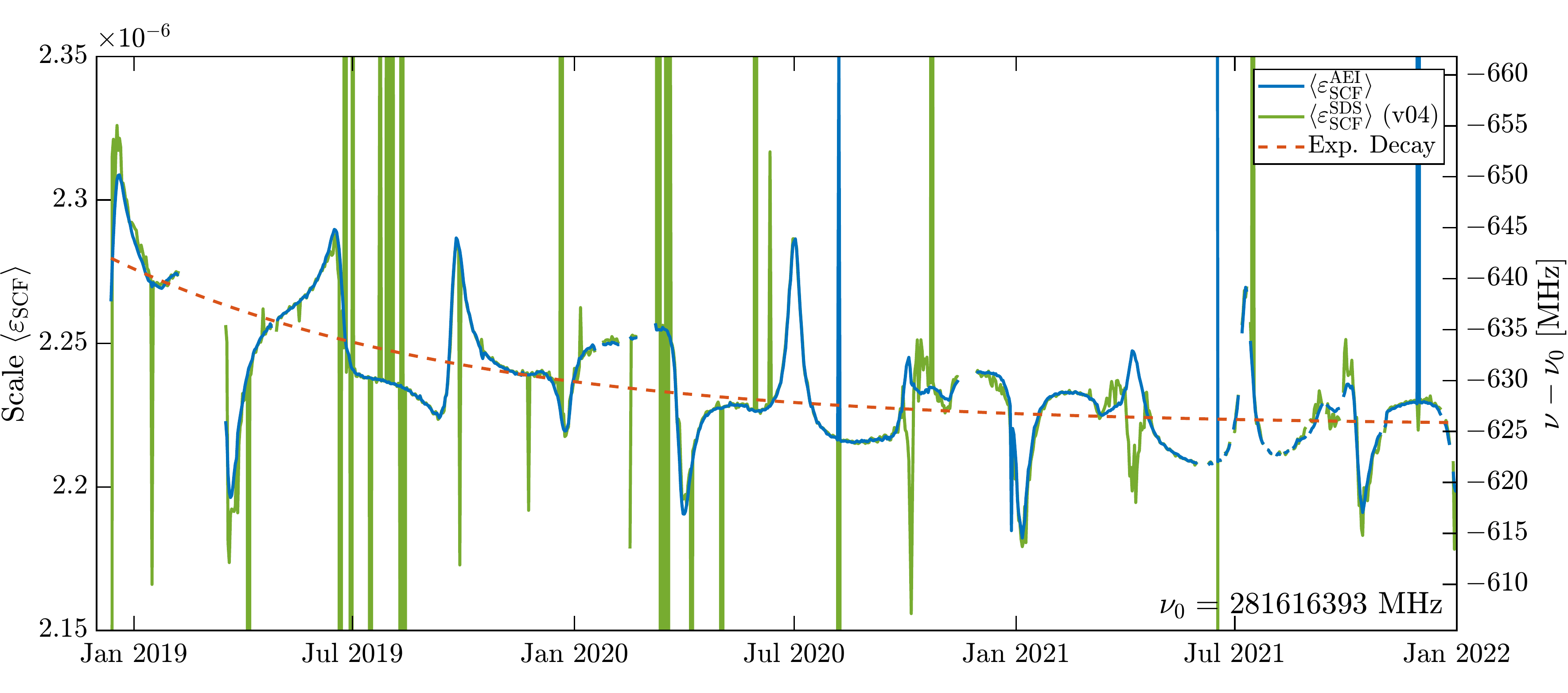}
	\caption{Comparison of LRI scale factor $\langle\epsSCF\rangle$ using the conventional cross-calibration method. \emph{Blue:} using the AEI ranging phase $\varphi_\mathrm{LRI}$, cf. \cref{eq::rawRange1}. \emph{Green:} the SDS LRI1B-v04 result. \emph{Orange dashed:} Exponential model for the cavity resonance frequency.}
	\label{fig::scales_SDS}
\end{figure}
The minimization result for $\langle\epsSCF^\mathrm{SDS}\rangle$, as given in the LRI1B-v04 data product, is shown in green in \cref{fig::scales_SDS}. Our recomputation with an in-house Level 1A to 1B processing is denoted as $\langle\epsSCF^\mathrm{AEI}\rangle$ (blue), which will be used later on as one possible frequency model. The plot covers the timespan from 2018-Dec-13 until 2022-Jan-01, where \GFC acts as the LRI reference unit. Due to spacecraft-related outages, the LRI was not in science mode from 2019-Feb-06 to 2019-Mar-17. Smaller gaps in the data originate from phase breaks, e.\,g., due to spacecraft maneuvers or diagnostic data recording. The frequent data gaps starting in mid-2021 are due to nadir-pointing of the spacecraft, occurring roughly two days per week. In these periods, the pointing angles between the spacecraft-fixed coordinate system and the line-of-sight exceed the \gls{LRI} pointing capabilities.

Even though the blue and green traces roughly match, the SDS implementation seems less robust as it shows more outliers, which may be related to imperfect phase jump removal \cite{Abich2019}. The number of outliers reduced after 2020-Jun-27, when the deglitching algorithm was adjusted by the \gls{SDS} \cite{L1ReleaseNotes}.
Both traces show a slow drift that seems to converge and peaks and dips occurring roughly every three months, indicating an apparent change of the laser frequency with a magnitude of \num{\pm e-7} or \SI{\pm 20}{\mega\hertz}. It is noteworthy that we can not distinguish which instrument contributes to those periodic variations, as we always use the difference between \gls{LRI} and \gls{KBR}. However, \cref{sec::inflightapplication,sec::toneErrors} will address these variations in more detail. 
The slow drift in \cref{fig::scales_SDS} was fitted as exponential decay of the form 
\begin{equation}
	\eps^\mathrm{Cav}(t) = \eps^\mathrm{Cav}_\infty - \eps^\mathrm{Cav}_0 \exp(-\lambda^\mathrm{Cav}\cdot t) \label{eq::epsCav}
\end{equation}
and is shown in orange. The decay rate is $\lambda^\mathrm{Cav} = \cavityDriftDecayRate$, with $\eps_\infty^\mathrm{Cav} = \cavityDriftAsymptote$ and $\eps_0^\mathrm{Cav}=\cavityDriftInitial$, the time $t$ is \gls{GPS} seconds past 2018-May-22, 00:00:00 UTC. 
Exponential shrinkage (and thus increasing frequency) has already been observed in similar cavities made from \gls{ULE} materials, and the suspected cause is aging of the spacer material \cite{Alnis2008}. 
\Cref{eq::epsCav} can of course be converted into an equivalent frequency model $\nu_\RR^\mathrm{Cav}$ via \cref{eq::nuEps}. This exponential model is the second model for the laser frequency, resulting in similar values as the \gls{SDS} scheme, but without the periodic features. 
Until now, we only derived these exponential model coefficients for \GFC. The derivation for \GFD is beyond the scope of this manuscript as \GFD was only for short times in reference mode.

The scheme of cross-calibration is only possible due to the unique situation of having two independent ranging measurements by \gls{KBR} and \gls{LRI}. However, it can not resolve intraday frequency variations of the \gls{LRI} and introduces small discontinuities at day bounds. Furthermore, it depends on the \gls{KBR}, which will likely not be present in future missions. 
Therefore we present an on-ground calibration that has been performed for the two laser flight models as a third method to determine the laser frequency and derive a calibrated frequency model only using telemetry data of the \gls{LRI}. It is based on the fact that the laser frequency $\nu_\RR$ can be deduced from the setpoints of the laser's frequency-lock control loop and thermal state.
With this model, we can continuously evaluate the optical frequency in orbit with moderate accuracy. The laser frequency may change due to varying environmental conditions, e.\,g., temperatures of the optical reference cavity. This model will be derived in \cref{sec::lasertelemetry,sec::ongroundCalibration,sec::empiricalModel}, and all models are analyzed and compared to each other in \cref{sec::inflightapplication}.

\section{LRI Laser and Telemetry Description}
\label{sec::lasertelemetry}
The \gls{LRI} \glspl{RLU} were built by Tesat Spacecom and are comparable to the laser onboard LISA Pathfinder and to the seed laser of one possible LISA implementation \cite{LPF_collaboration_2017}. They are based on an Nd:YAG \gls{NPRO} crystal and are fiber-connected to an optical reference cavity built by Ball Aerospace \cite{Pierce_2012,Thompson2011} and to the \gls{OBA}. The laser's output power is in the order of \SI{25}{\milli\watt} in the near-infrared regime ($\lambda\approx\SI{1064}{\nano\meter}$) \cite{Nicklaus2017}. The laser frequency is actively controlled by feedback control loops either from the reference cavity using the \gls{PDH} scheme \cite{Drever1983} (in reference mode) or to the incoming beam using a frequency-offset lock (transponder mode). The tuning is achieved through a thermal element for slow variations and a \gls{PZT} actuator for fast variations. The actuator signals are downlinked in the laser telemetry and published within the LHK1A/B data products 
at a rate of \SI{1}{\hertz}, if the LRI is in science mode, i.\,e., when the laser link is established. The data type is unsigned integer of 32 bits depth. The corresponding normalized signed data streams are computed via the two's complement and scaling by the bit depth as
\begin{align}
	\texttt{u2i}(x, N) = &\left\{ \begin{matrix}
	x/2^N-1, & \mathrm{if~} x \geq 2^{N-1} \\
	x/2^N, & \mathrm{if~} x < 2^{N-1}
	\end{matrix}\right. \label{eq::f2c}
\end{align}
with $N=32$ and $x$ denoting the unsigned value from the telemetry. The value range is \(-1/2 < \texttt{u2i}(x,\cdot) \leq 1/2\).
In the following, these normalized data streams are denoted as \pztin, \pztout, \thermalin and \thermalout. The temperature of the laser can be retrieved from so-called ``OFFRED'' data, which is recorded by the \gls{OBC}. The measurement is taken at the \gls{TRP} of the \gls{RLU}, which is located at the laser's housing. By the time of writing, the laser \gls{TRP} temperature is not publicly available. Still, it will be shown later that the influence of the \gls{TRP} coupling is small during nominal operations.

The notations \gls{IL} and \gls{OOL} are not referring to different sensors as in conventional feedback control circuits, but two parts that are added to form the final setpoint. The OOL channel is used for manual control with some logic (e.\,g., to drive a frequency ramp for locking to the cavity or during acquisition). In contrast, the IL value represents the evolution of the actuator value in closed-loop operation.
The actuator range of the \gls{PZT} and thermal actuators is \SI{\pm1}{\volt} and \SI{\pm9}{\volt}, respectively, with nominal frequency coupling coefficients of \SI{5}{\mega\hertz\per\volt} and \SI{500}{\mega\hertz\per\volt}, respectively, also shown in \cref{tab::designvalues}.

\section{RLU On-Ground Calibration} \label{sec::ongroundCalibration}
\begin{figure}
    \centering
    \includegraphics[width=0.75\linewidth]{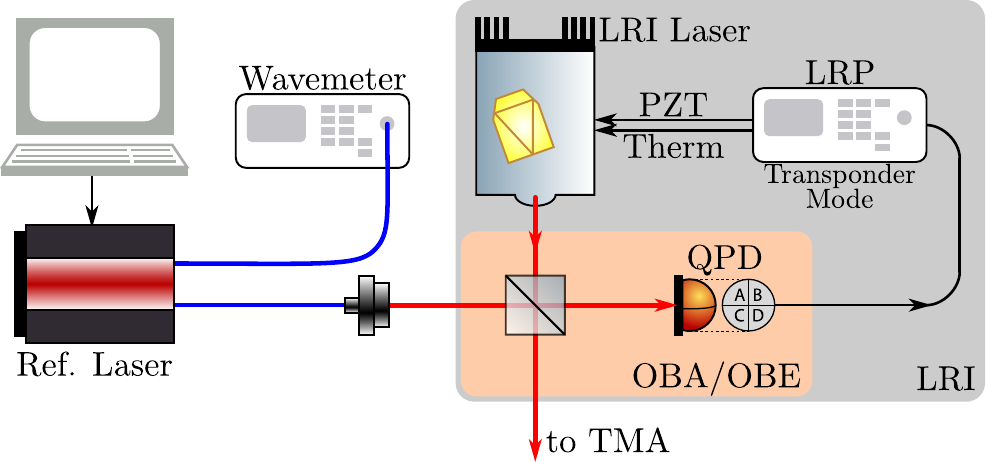}
    \caption{Laboratory setup for the \gls{LRI} flight laser frequency calibration measurement. Blue lines denote optical fibers; red lines are laser beams in free space. Black arrows denote electric signals.}
    \label{fig::expSetup}
\end{figure}

The laboratory setup used to calibrate the \gls{LRI} lasers is shown in \cref{fig::expSetup}. It consisted of the \gls{LRI} flight laser, the \gls{LRP} (including the phasemeter), a frequency-controlled reference laser and a wavemeter. The measurements were performed by parts of the \gls{LRI} teams at JPL/NASA and AEI Hannover.
During these activities, the frequency of the reference laser was tuned using a computer, and its frequency was recorded using a wavemeter. The \gls{LRI} unit in transponder mode locks its laser frequency to the incoming beam and adds a \SI{10}{\mega\hertz} offset and is thus known as well. During the activities, the RLU temperature, as well as the PZT and thermal telemetry, is recorded. The frequency of the \gls{LRI} laser was not measured directly, since it was more convenient to use the second output port of the reference laser (one fiber to the wavemeter, one to the optical bench), while the \gls{LRI} laser light is free-beam on the optical bench. 
We use a linear model to estimate the laser frequency based on the \gls{TM}:
\begin{equation}
	\nu^\mathrm{TM}(t) = \begin{pmatrix}
		c_\pztin     \\ c_\pztout \\
		c_\thermalin \\ c_\thermalout \\
		c_\lastrp
	\end{pmatrix} \cdot
	\begin{pmatrix}
		\pztin(t)     \\ \pztout(t) \\
		\thermalin(t) \\ \thermalout(t) \\
		\lastrp(t-\tau) - \SI{26}{\degreeCelsius}
	\end{pmatrix} 
	+ \nu_{0,\mathrm{air}} + \Delta\nu_\mathrm{AirToVac}\ ,
	\label{eq::lsq_ansatz}
\end{equation}
which depends on the actuator states, i.\,e., the telemetry data streams \pztin, \pztout, \thermalin, \thermalout as well as the surrounding temperature, which is measured at the \gls{TRP} of the laser. Since the \gls{TRP} is located outside the thermal shielding, a time delay of $\tau=\SI{520}{\second}$ is applied to the temperature measurements, which represents the propagation time of outer temperature changes to the \gls{NPRO} crystal. The manufacturer provided this numerical value. Furthermore, only deviations from the nominal temperature of \SI{26}{\degreeCelsius} are considered.  

\begin{figure}
	\centering
	\includegraphics[width=\linewidth]{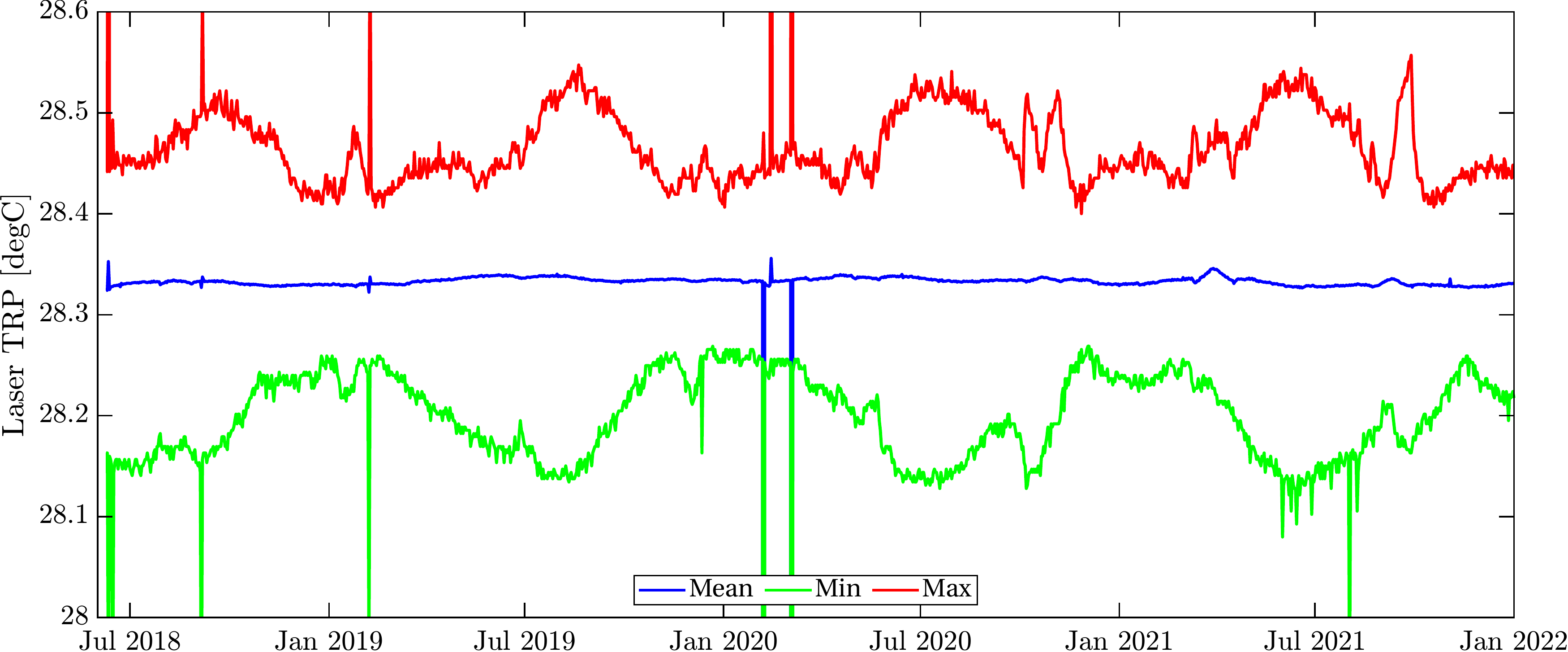}
	\caption{Daily mean, min and max of the \GFC \lastrp data.}
	\label{fig::lasTRP}
\end{figure}

The nominal values for the coupling factors given by the laser manufacturer are shown in \cref{tab::designvalues}. However, we refine the individual laser units' thermal coupling coefficients with our measurements. The \gls{PZT} and \gls{TRP} coupling are not refined since they were not modulated strongly enough during the calibration measurements to derive reliable coupling factors. 
We expect the TRP coupling to be non-critical since the lasers' \gls{TRP} temperature varies only in the sub-Kelvin domain in flight as shown in \cref{fig::lasTRP}. The blue trace depicts the daily averaged laser TRP recording of \GFC and its respective daily minimum (green) and maximum (red) values. 
The temperature of the \GFC laser is stable when averaged daily. It shows sub-daily variations of $\pm\SI{0.2}{\kelvin}$, which translates into $\Delta\nu=\pm\SI{2.4}{\mega\hertz}$ in frequency, or $\epsSCF \approx \Delta\nu/\nu \approx \num{\pm 8.5e-9}$ using a coupling of \SI{-12}{\mega\hertz\per\kelvin} (cf. \cref{tab::designvalues}).
\begin{table}[p]
	\centering
	\caption{Coupling factors for the two LRI laser flight units. Shown are the design values provided by the laser manufacturer and fit results from on-ground measurements. \gls{PZT} and \gls{TRP} coupling factors were not refined, since the measurements were not suitable to derive these couplings. The static value $\Delta\nu_\mathrm{AirToVac}$ was provided by the manufacturer and represents the frequency change from air to vacuum. The shown uncertainties are the formal errors of the least squares estimation.}
	\label{tab::designvalues}
	\begin{tabularx}{\textwidth}{*{3}{c}*{2}{Y}}
		\toprule
		Coupling                      & Unit    & Design value                                                               & \GFC (fit)       & \GFD (fit)      \\
		\midrule
		$c_\pztin$                    & [MHz]   & $\SI{2}{\volt}\cdot\SI{5}{\mega\hertz\per\volt}=\SI{10}{\mega\hertz}$      & -          & -          \\
		$c_\pztout$                   & [MHz]   & $\SI{18}{\volt}\cdot\SI{5}{\mega\hertz\per\volt}=\SI{90}{\mega\hertz}$     & -          & -          \\
		$c_\thermalin$                & [MHz]   & $\SI{2}{\volt}\cdot\SI{500}{\mega\hertz\per\volt}=\SI{1000}{\mega\hertz}$  & $\num{1097}\pm\num{0.383}$ & $\num{1094}\pm\num{1.383}$ \\
		$c_\thermalout$               & [MHz]   & $\SI{18}{\volt}\cdot\SI{500}{\mega\hertz\per\volt}=\SI{9000}{\mega\hertz}$ & $\num{9155}\pm\num{7.396}$ & $\num{8857}\pm\num{27.565}$ \\
		$c_\lastrp$                   & [MHz/K] & \num{-12}                                                                  & -          & -          \\
		$\nu_{0,\mathrm{air}}$        & [MHz]   & \makecell{\num{281614803} (\GFC)                                                                                    \\ \num{281614780} (\GFD)} & \makecell{$\num{281614682.081}$\\$\pm\num{0.378}$} & \makecell{$\num{281614631.999}$\\$\pm\num{1.094}$} \\
		$\Delta\nu_\mathrm{AirToVac}$ & [MHz]   & \makecell{\num{37} (\GFC)                                                                                           \\ \num{27} (\GFD)} & - & - \\
		\bottomrule
	\end{tabularx}
\end{table}
\begin{figure}[p]
	\def\figwidth{0.48\linewidth}
	\centering
	\subfloat[PZT control loop signals]{
		\includegraphics[width=\figwidth]{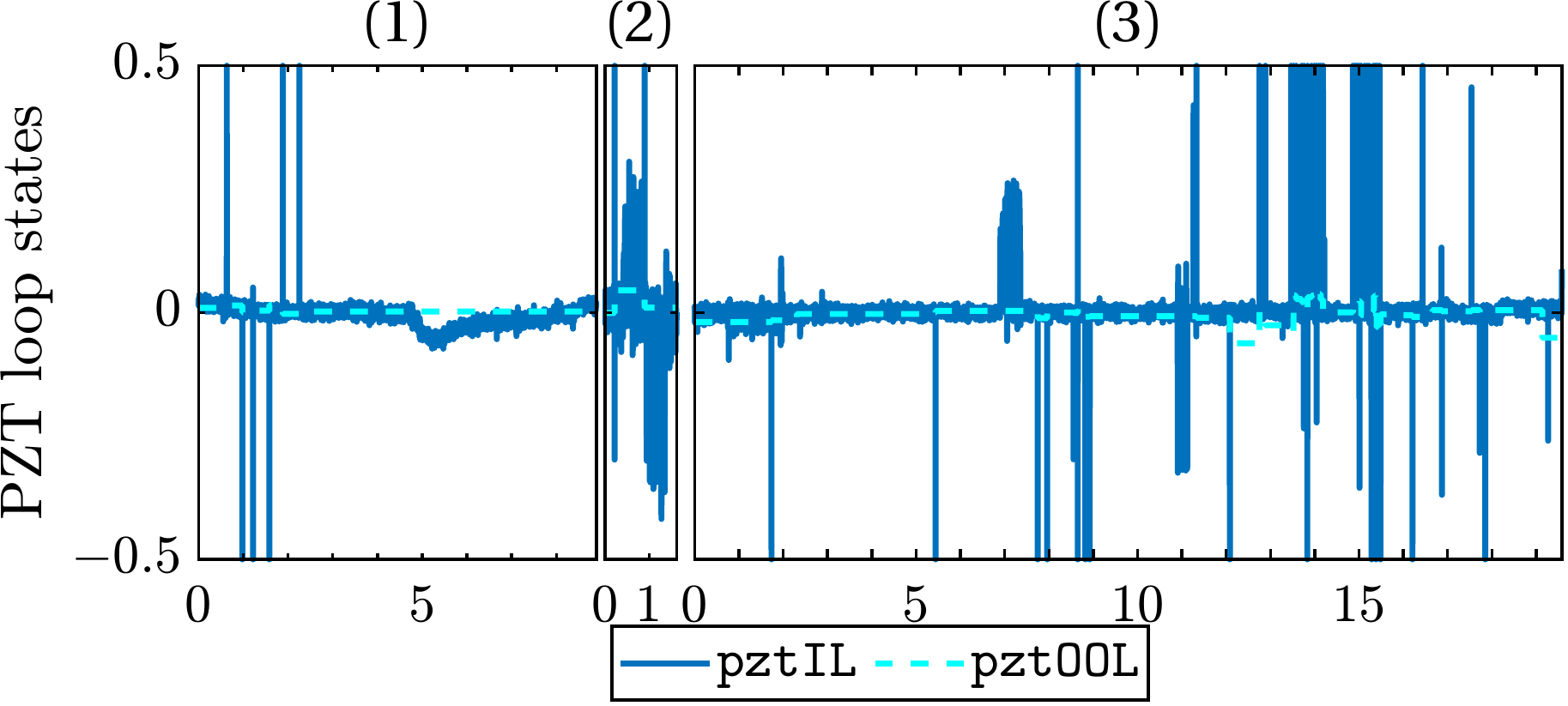}
	}
	\subfloat[Thermal control loop signals]{
		\includegraphics[width=\figwidth]{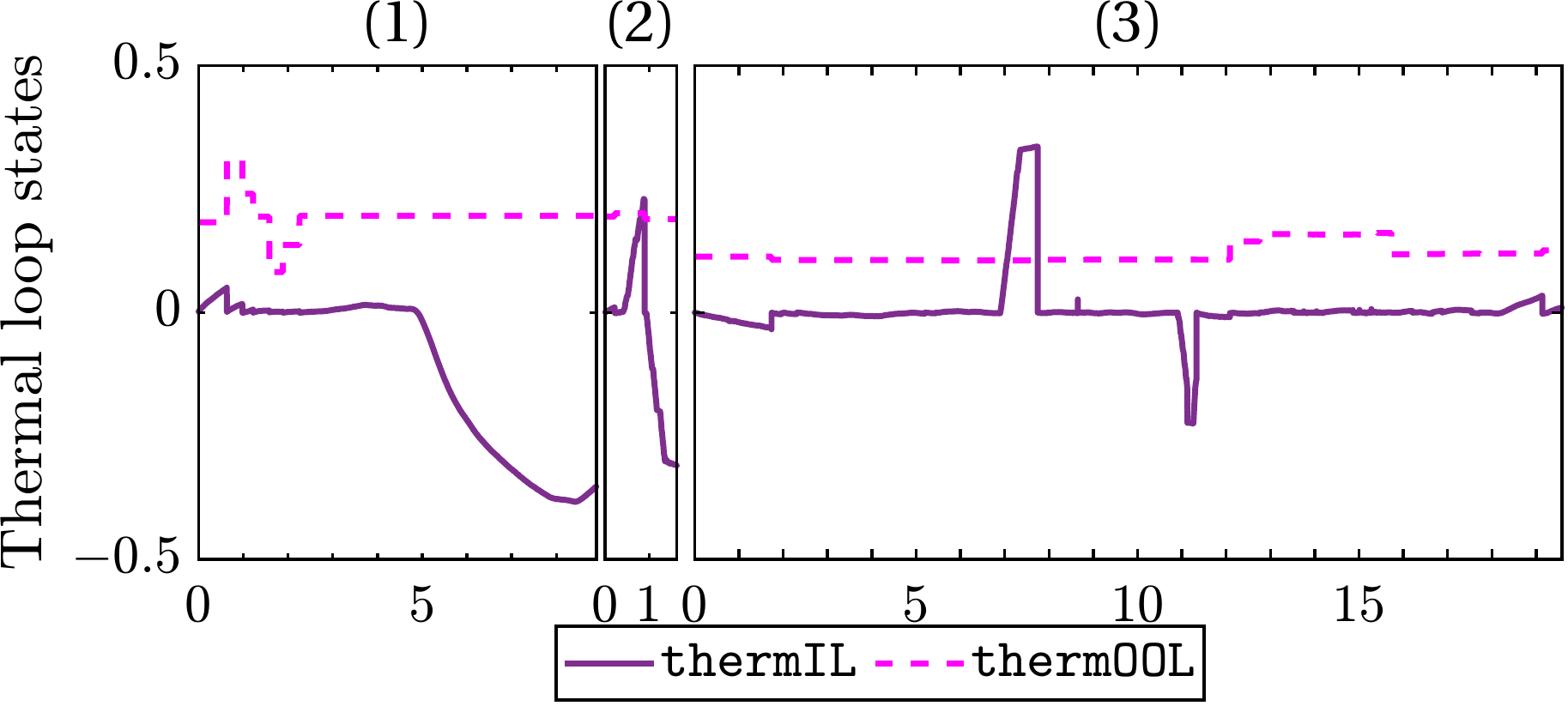}
	}
	\hfil
	\subfloat[Temperature at the laser TRP.]{
		\includegraphics[width=\figwidth]{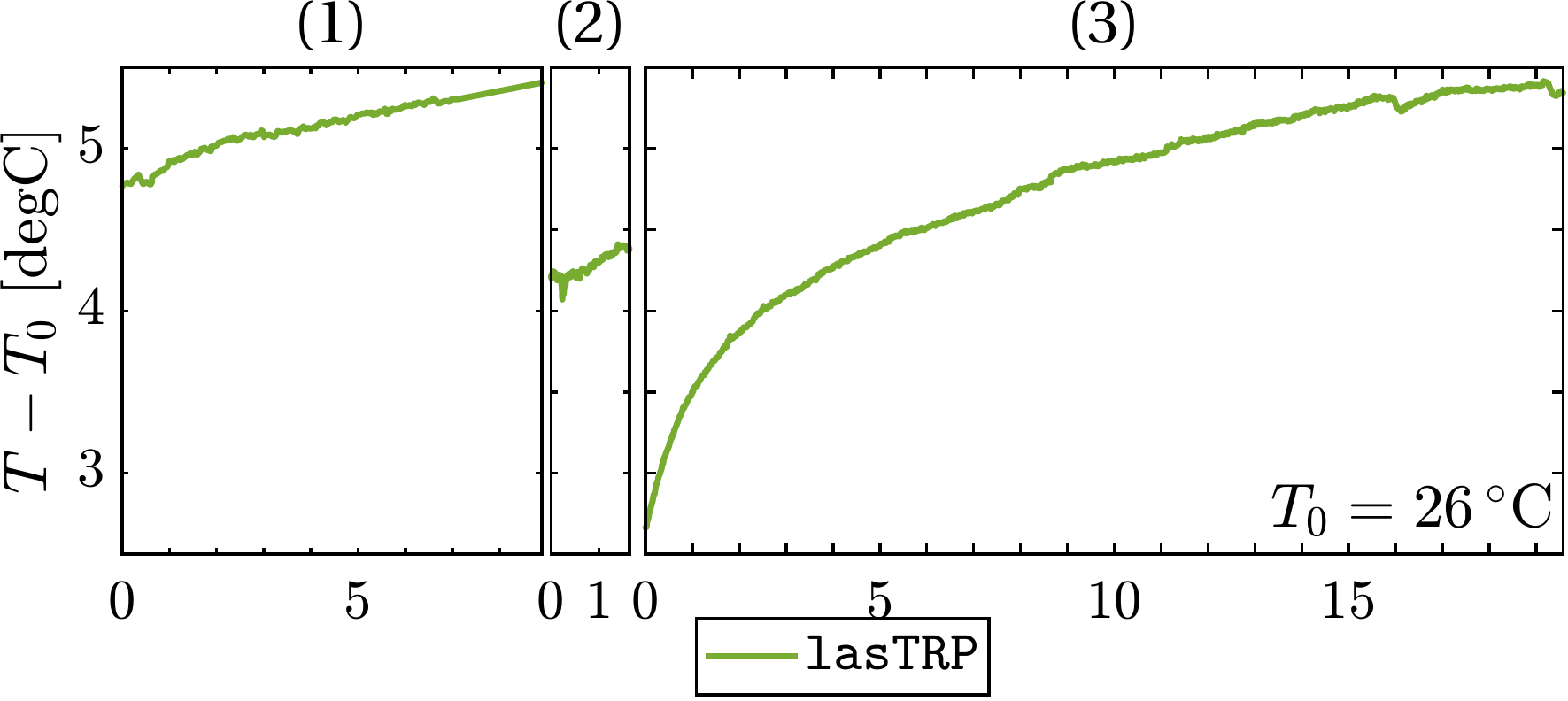}
	}
	\subfloat[Measured frequency $\nu^\mathrm{ref}$ and model $\nu^\mathrm{TM}$.]{
		\includegraphics[width=\figwidth]{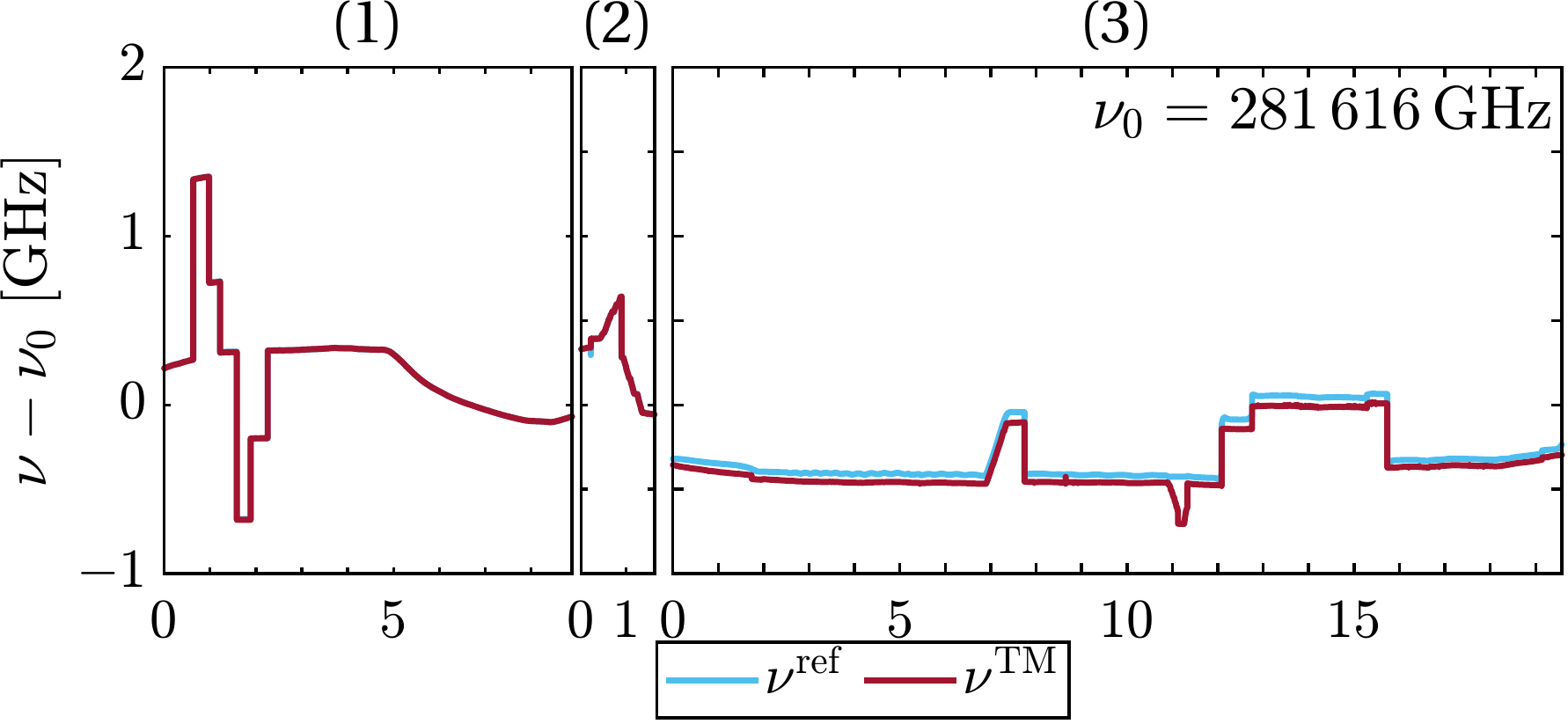}
	}
	\hfil
	\subfloat[Frequency residuals $\nu^\mathrm{ref}-\nu^\mathrm{TM}$]{
		\includegraphics[width=\figwidth]{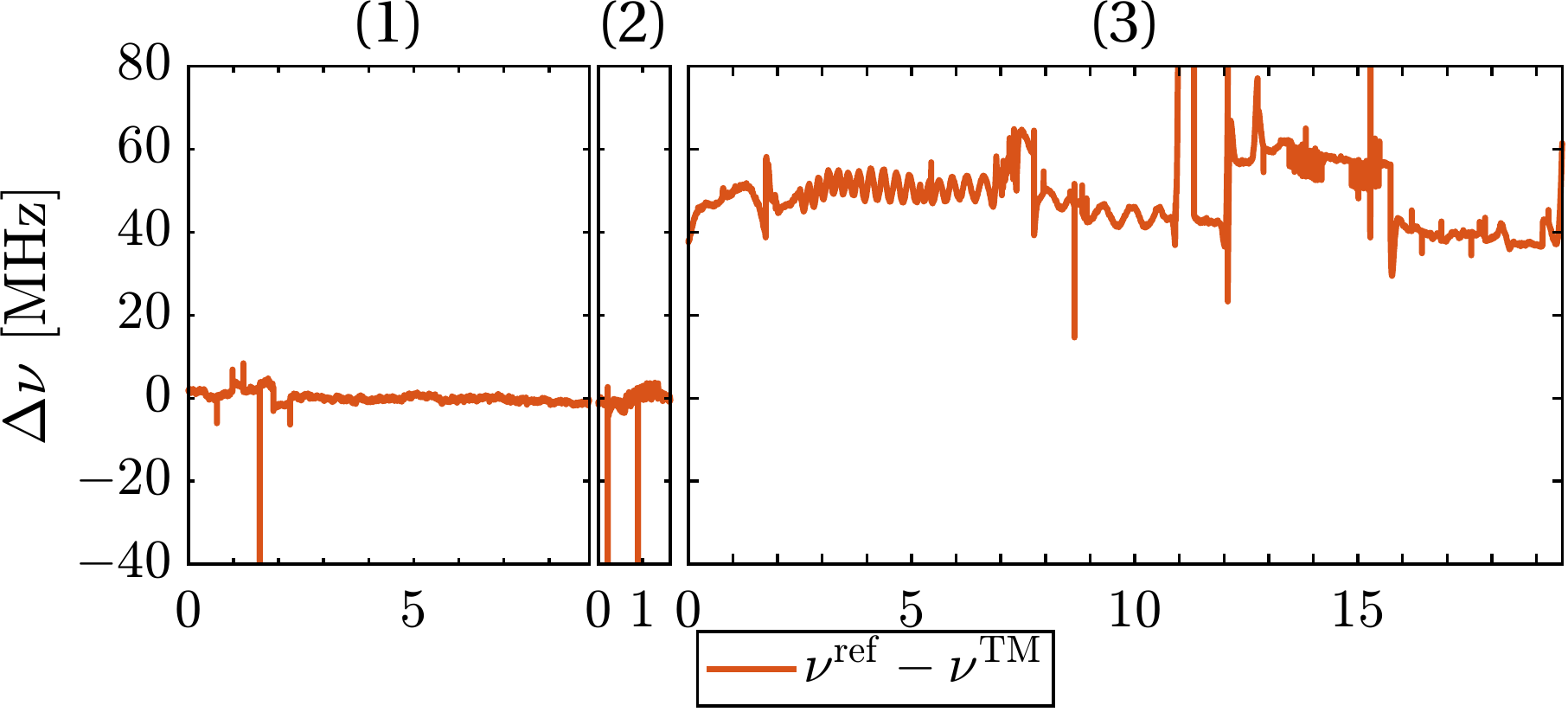}
	}

	\caption{Regression results for the \GFD laser. The numbering on top of the individual panels of each subfigure corresponds to the measurement campaigns, as explained in the text. Note the offset in the residuals in (e) when the less precise wavemeter WS6-600 was used in campaign (3). The average bias here is \SI{51.494\pm0.24}{\mega\hertz}. All x-axes are in arbitrary time units.}
	\label{fig::RLU27_results}
\end{figure}

Several calibration measurements were performed on both \glspl{RLU} between July 2017 and January 2018. For the laser integrated into \GFD, four measurements were taken. In the following, we label these four measurements (1)...(4). They all differ a little in their procedure. In (1) the reference laser's frequency was commanded in discrete steps, which caused the \gls{LRP} to lose lock and forced reacquisition and thus a temporary data loss. Afterward, the reference laser was put into a free-running cool-down mode without active stabilization. Reacquisition was avoided in (2) by sweeping continuously over the same frequency range. (3) was a diagnostic test for the \gls{DWS}, and the absolute frequency measurement was a secondary result. Test (4) consists of very few sample points only since the used wavemeter had no digital output port but only a display to retrieve the data. Thus, this analysis does not use the data of (4). The measurements (1) and (2) were performed in July 2017 using a HighFinesse WS7-60 wavemeter with an absolute accuracy of \SI{60}{\mega\hertz}. Test (3) in November 2017 used a HighFinesse WS6-600 (\SI{600}{\mega\hertz} accuracy) and in (4), a Burleigh WA1500 (\SI{60}{\mega\hertz} accuracy) was used. The \GFC laser was tested twice: once with a WS6-600 in November 2017 and a Burleigh WA1500 in January 2018, and again, the latter one is not used in this analysis. 

We use a least squares approach to estimate the linear coupling factors and constants of \cref{eq::lsq_ansatz}. Additionally, we weigh the WS7-60 measurements higher by a factor of 5 compared to the WS6-600, which has lower accuracy. We furthermore estimate a relative offset of the WS6-600 wavemeter, which we can deduce by analyzing the residuals. This offset of the WS6-600 is also apparent when measuring an absolute frequency reference like an iodine cell; see \cref{app::ws6-iodine,app::ws7-iodine} for more information.

\Cref{fig::RLU27_results} shows the regression result using the measurements for \GFD. The individual measurement campaigns are labeled (1)...(3). The subfigures (a) and (b) show the normalized telemetry of the laser control loops, and the temperature of the laser's \gls{TRP} is shown in subfigure (c). Panel (d) contains the absolute frequency of the reference laser $\nu^\mathrm{ref}$ (shifted by \SI{10}{\mega\hertz} to compensate for the offset-frequency lock of the \gls{LRP}) and the resulting laser frequency model $\nu^\mathrm{TM}$ of the \gls{LRI} laser. The trace in (e) shows the residuals $\nu^\mathrm{ref}-\nu^\mathrm{TM}$, which clearly exhibits an offset of approximately \SI{50}{\mega\hertz} beginning at (3), where the WS6-600 was used. The high-frequency variations are higher in (3) due to the lower precision of the WS6-600. The resulting coupling factors from the linear least squares minimization are shown in \cref{tab::designvalues}. Generally, the resulting values match the manufacturer's design values with only slight deviations.

\section{Empirical Refinement of Telemetry-Based Laser Frequency Model}
\label{sec::empiricalModel}
\begin{figure}[tb]
	\centering
	\includegraphics[width=\linewidth]{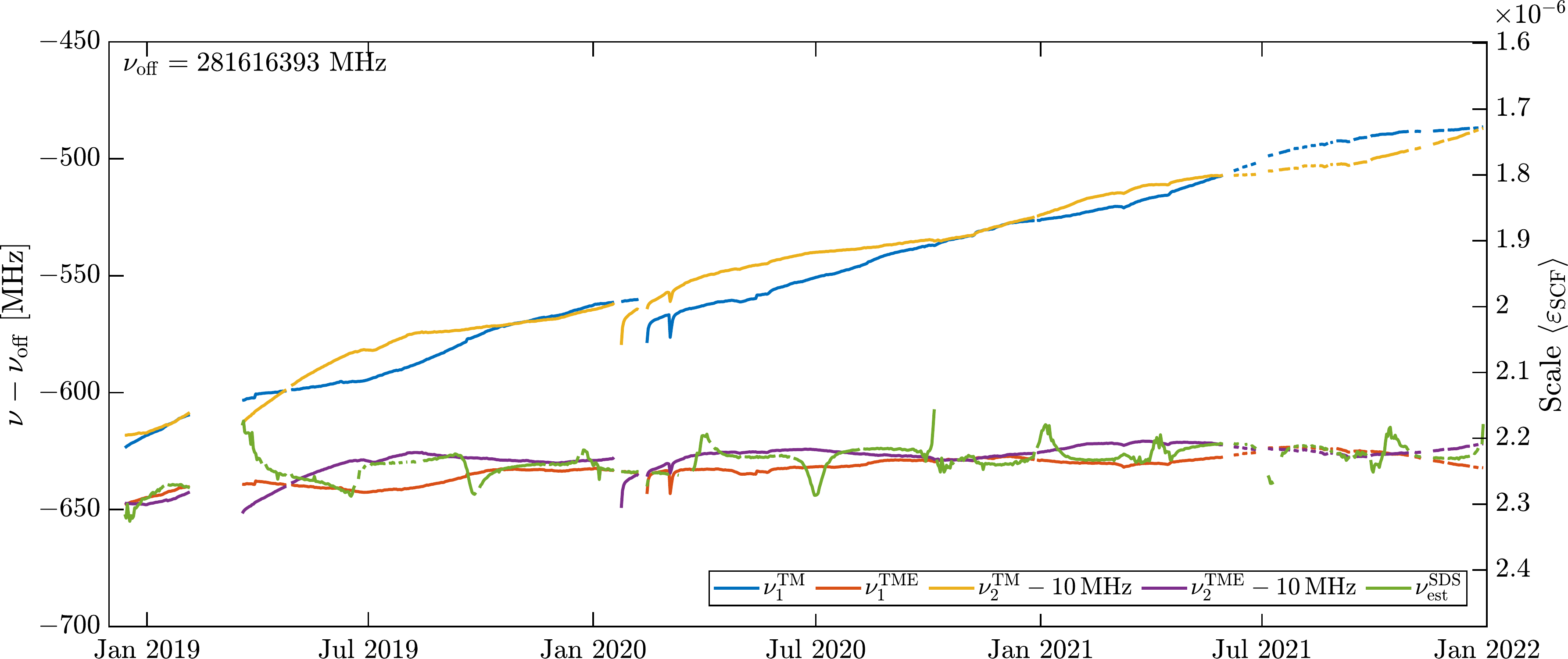}
	\caption{Purely ground calibration-based models $\nu^\mathrm{TM}_{1/2}$ and empirically corrected \gls{TM} models $\nu^\mathrm{TME}_{1/2}$ for \GFC and \GFD laser frequencies alongside the \gls{SDS} frequency $\nu^\mathrm{SDS}_\mathrm{est}$ from KBR-LRI cross-calibration. Outliers in SDS curve removed. The right axis shows approximate equivalent laser frequency variations, cf. \cref{eq::nuEps}.}
	\label{fig::emp_fit_results}
\end{figure}
The calibrated telemetry models $\nu^\mathrm{TM}_{1/2}$ are now compared to the frequency $\nu_\mathrm{est}^\mathrm{SDS}$ (cf. \cref{eq::nuEps}) from the KBR-LRI cross-correlation, where the flight data spanning from 2018-Dec-13 until 2022-Jan-01 is used. \Cref{fig::emp_fit_results} shows the frequency estimates from the \gls{TM} models for both spacecraft (blue and yellow) alongside the \gls{SDS} frequency (green). The latter is already shown in \cref{fig::scales_SDS}, but outliers were removed this time. The \GFD curves are shifted down by \SI{10}{\mega\hertz} to remove the intended transponder frequency offset (cf. \cref{sec::theory}). The subscript $1/2$ refers to \GFC or \GFD, respectively.

The models (blue and yellow) differ by \SI{20}{\mega\hertz} at maximum, which is within the accuracy of the better wavemeter WS7-60, defining the model accuracy. However, a drift of the models w.\,r.\,t. the KBR cross-calibration method (green) is visible. The current hypothesis to explain this drift is an aging effect of the \gls{NPRO} crystal or the electronics within the LRP. However, there is little literature on aging-induced frequency changes of \gls{NPRO} lasers, and this theory might need verification in a laboratory experiment. The drift appears only in the laser setpoint telemetry but not in the frequency, which is tightly locked to the cavity resonance.

The curves show some data gaps starting in mid-2021, caused by regular nadir-pointing periods, in which the \gls{LRI} was not creating science data.
The steep slopes and the dip in February and March 2020 in $\nu^\mathrm{TM}_{\,1/2}$ are due to spacecraft-related non-science phases of the LRI, after which the units had to heat up to reach the nominal temperatures. This heating process is visible at the laser \gls{TRP} (cf. \cref{fig::lasTRP}) and thus affects not only the TM model but also the green \gls{SDS} curve with comparable magnitude, which confirms the temperature coupling estimate in the TM model.
However, we found that the link acquisition happened before the lasers reached their thermal equilibrium, which led to an apparent small step in the $\nu_{1/2}^\mathrm{TM}$ frequency model (see \cref{app::thermIlOol}). Imperfect coupling factors could possibly cause this. 
To account for these steps $\nu^\mathrm{step}_i$ in our telemetry-based laser frequency model, as well as for the drifts $a$ and offsets $\Delta\nu$ from the \gls{NPRO} aging, we define an empirical correction and estimate its parameters by least squares minimization using $\nu_\mathrm{est}^\mathrm{SDS}$ as the reference. The empirical model reads
\begin{equation} \label{eq::driftModel}
	\nu^\mathrm{emp}(t) = a\cdot(t-t_0)+\Delta\nu\ +\,\nu^\mathrm{step}(t)\ ,
\end{equation}
where the reference epoch $t_0 = 1210982400$ \gls{GPS} is 2018-May-22 at midnight. The steps are defined as 
\begin{align}
	\nu^\mathrm{step}(t) = \nu_i^\mathrm{step} \mathrm{~~if~~} t^\mathrm{step}_i \leq t < t^\mathrm{step}_{i+1}\ .
\end{align}
The estimated parameters $a$ and $\Delta\nu$ are shown in \cref{tab::FrequencyDriftParameters}, while the steps $\nu^\mathrm{step}_i$ and the corresponding time-tags $t^\mathrm{step}_i$ are shown in \cref{tab::FrequencyStepParameters}. Unfortunately, this empirical model makes the telemetry-based frequency model still dependent on the \gls{KBR}. In principle, one could overcome the needs of an empirical model by better calibrating the laser before launch.
\begin{table}[tb]
	\centering
	\caption{Parameters for the empirical part of the laser frequency model for \GFC and \GFD of \cref{eq::driftModel}.}
	\label{tab::FrequencyParameters}
	\subfloat[Drift Parameters\\~]{
		\begin{tabularx}{.45\linewidth}{*{4}{Y}}
			\toprule
			Coupling        & Unit              & \multicolumn{2}{c}{Value}  \\
			& & \GFC & \GFD \\
			\midrule
			$a$  & [\si{\hertz\per\second}] & \num{1.419}  & \num{1.110} \\
			$\Delta\nu$ & [\si{\mega\hertz}]    & \num{-5.881} & \num{-1.089} \\
			\bottomrule
		\end{tabularx}
		\label{tab::FrequencyDriftParameters}
	}
	\hfil
	\subfloat[Steps in the telemetry-based laser frequency model. Time tags refer to midnight.]{
		\begin{tabularx}{.5\linewidth}{c*{3}{Y}}
			\toprule
			$i$ & $t^\mathrm{step}_i$ & \multicolumn{2}{c}{$\nu^\mathrm{step}_i$ [MHz]} \\
			& & \GFC & \GFD \\
			\midrule
			1 & 2018-05-22 & 0 & \num{0} \\
			2 & 2020-01-10 & 0 & \num{-1.770} \\
			3 & 2020-02-10 & \num{-11.593} & \num{2.881} \\
			4 & 2020-05-11 & \num{-10.446} & \num{2.881} \\
			5 & 2022-01-01 & undefined & undefined \\
			\bottomrule
		\end{tabularx}
		\label{tab::FrequencyStepParameters}
	}
\end{table}
This empirical model is subtracted from the telemetry model to form the final telemetry-based and empirically corrected TME frequency estimate \glsreset{TM}
\begin{equation}
    \nu_{1/2}^\mathrm{TME}(t) = \nu^\mathrm{TM}_{1/2}(t) - \nu^\mathrm{emp}_{1/2}(t)\ . \label{eq::nu12}
\end{equation} 
After applying the empirical model, the numerical values of the total frequency models $\nu^\mathrm{TME}_{1/2}$ for \GFC and \GFD (orange and purple traces in \cref{fig::emp_fit_results}) are in the range of $\nu^\mathrm{SDS}_\mathrm{est}$. Besides the cross-calibration method, the telemetry-based model does not show seasonal or periodic features. Note that the exponential drift of the cavity is contained in $\nu^\mathrm{SDS}_\mathrm{est}$ and $\nu_{1/2}^\mathrm{TME}$, even though it is hard to see in the \cref{fig::emp_fit_results}. However, the empirical model in \cref{eq::driftModel} does not absorb the effect of exponentially increasing frequency (cf. \cref{fig::scales_SDS,eq::epsCav}), since that cavity drift is present in both, $\nu^\mathrm{TME}_1$ and the reference $\nu^\mathrm{SDS}_\mathrm{est}$, thus it is not apparent in the metric of the least squares adjustment.

\section{Comparison of the LRI1B-Equivalent Data Sets} \label{sec::inflightapplication}
We define the pre-fit range error based on the instantaneous range difference between the \gls{LRI} and \gls{KBR}, 
\begin{equation}
	\rho_\mathrm{err,v5X}^\mathrm{pre}(t) = \rho_\mathrm{LRI,v5X}^\mathrm{inst}(t) - \rho_\mathrm{KBR}^\mathrm{inst}(t) - \rho_\mathrm{KBR}^\mathrm{FV}(t)\ .
	\label{eq::rhoErrPrefit_noFilt}
\end{equation}
The subscript v5X denotes three different versions of the LRI1B data product derived at the \gls{AEI} \cite{Mueller2021_MSc,Mueller2022}. They differ by the models for the laser frequency $\nuRG$. At first, the data product using the telemetry-based model described in the previous section (cf. \cref{eq::nu12}) is called LRI1B-v51. The exponential cavity model (cf. \cref{eq::epsCav}) forms LRI1B-v52. The last data product, LRI1B-v53, uses the predetermined, constant value $\nu_0$ only, which makes it, in principle, a pre-release of LRI1B-v04 without the daily scale $\langle\epsSCF^\mathrm{SDS}\rangle$ and timeshift $\zeta$ applied. The other differences between all three versions and the official v04 data are the improved deglitching algorithm \cite{Mueller2021_MSc} and the \gls{LTC} implementation according to Yan et al. \cite{Yan2021}. The LRI ranging data for these three versions at a \SI{10}{\hertz} data rate is derived for the time spanning from 2018-Dec-13 until 2022-Jan-01.
We further used a correction $\rho_\mathrm{KBR}^\mathrm{FV}$ for the intra-day carrier frequency variations of the \gls{KBR}. This correction mainly contains signal at 1/rev and 2/rev frequency \cite{Mueller2022} and improves the consistency between \gls{SDS}-derived \gls{KBR} and \gls{AEI}-derived \gls{LRI} data products, since the AEI-derived \gls{LRI} products inlude such a correction arising from the difference between proper time and coordinate \gls{GPS} time (cf. $\nu_\RR(t)$ vs. $\nu_\RR^\GG(t)$ in \cref{sec::errorCouplingModel}). However, the magnitude of this effect is small, and the results barely change when the correction is omitted.

\begin{figure}[p]
	\begin{adjustwidth}{-\extralength}{0cm}
	
		\subfloat[LRI1B-v51: Telemetry-based laser frequency model $\nu_{1/2}^\mathrm{TM}$]{
			\centering
			\includegraphics[width=.98\linewidth]{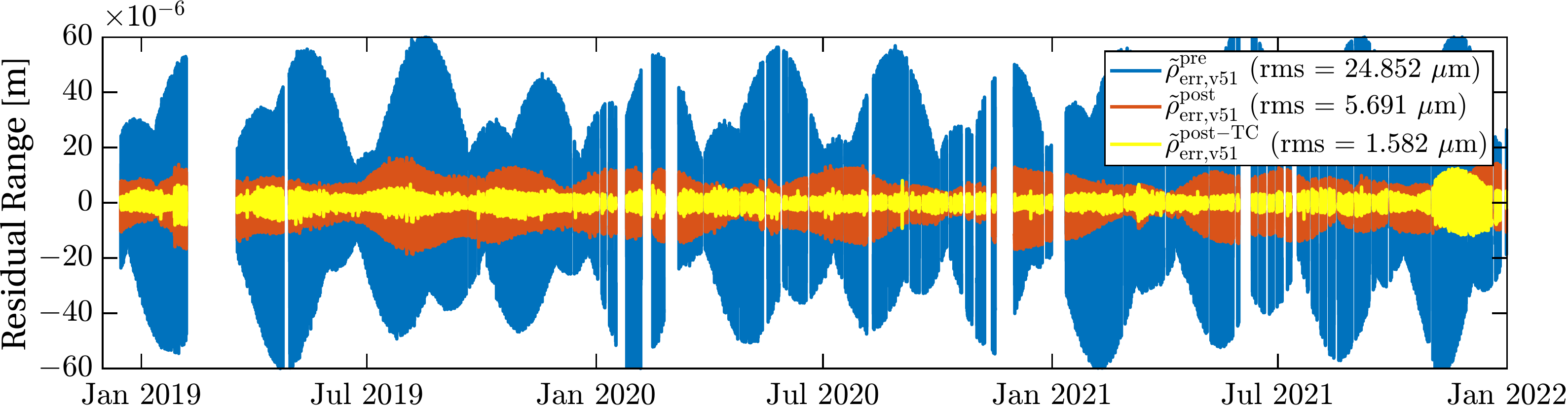}
		}
	
		\subfloat[LRI1B-v52: Exponential cavity frequency decay model $\nu_\RR^\mathrm{Cav}$]{
			\centering
			\includegraphics[width=.98\linewidth]{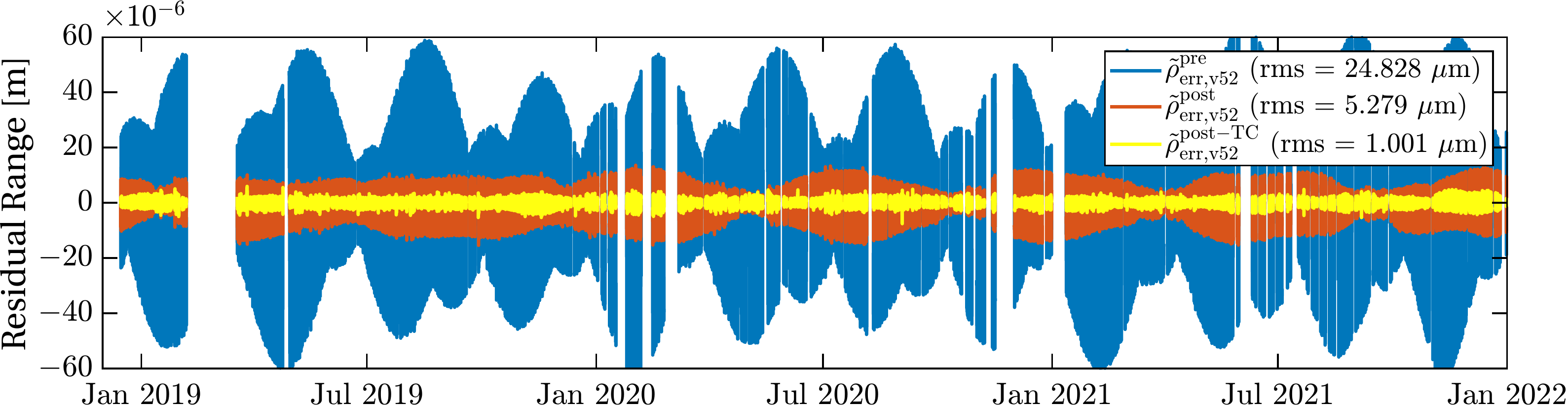}
		}
		
		\subfloat[LRI1B-v53: Pre-calibrated fixed frequency value $\nu_0$. The prefit difference KBR-LRI is large, since the constant frequency $\nu_0$ is a few \si{\mega\hertz} off from the truth.]{
			\centering
			\includegraphics[width=.98\linewidth]{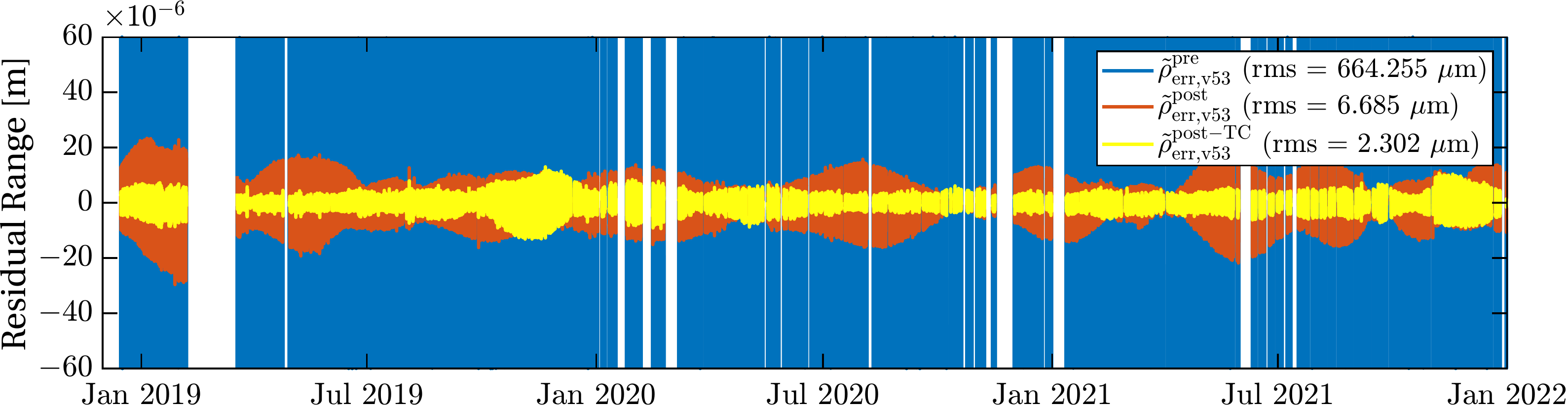}
			\label{fig::postfit_rhoErr_v53}
		}
	\end{adjustwidth}
	\caption{Prefit range error (\cref{eq::rhoErrPrefit}, blue) and postfit range error before (\cref{eq::rhoErrPostfit}, orange) and after (\cref{eq::rhoErrPostfitTC}, yellow) \glsentryshort{TC} fitting for all three frequency models. Initially, the prefit KBR-LRI range error shows rms variations of \SI{25}{\micro\meter}, \SI{25}{\micro\meter}, \SI{664}{\micro\meter} for v51, v52 and v53, respectively. For the orange traces, the effect of a global scale $\langle\epsSCF^\mathrm{glo}\rangle$ and timeshift $\zeta^\mathrm{glo}$ is subtracted, which already removes large parts of the residual signal (\SI{5.7}{\micro\meter}, \SI{5.3}{\micro\meter}, \SI{6.7}{\micro\meter}). After removal of the full \glsentryshort{TC} including five thermistors (yellow), the postfit range error is further reduced to a rms level of \SI{1.6}{\micro\meter}, \SI{1}{\micro\meter} and \SI{2.3}{\micro\meter}.}
	\label{fig::postfit_rhoErr}
\end{figure}

In general, the range error exhibits long-term drifts in the order of a few \SI{10}{\micro\meter\per day}, which we remove through a high-pass \gls{FIR} filter with a cutoff frequency of $\SI{0.08}{\milli\hertz}\approx f_\mathrm{orb}/2$. Future studies may address the reason for these long-term drifts, but this is beyond the scope of this article at the current stage. As mentioned in \cref{sec::SCFdetermination}, the \gls{LRI} scale factor is sensitive to variations or errors at 1/rev and 2/rev frequencies, which are unaffected by the filter. Due to the filtering, half a day of data is cropped at the start and end of each continuous segment, i.\,e., at every loss of the interferometric link of either \gls{KBR} or \gls{LRI}, to remove the initialization of the \gls{FIR} filter. Hence, all gaps appear longer than they actually are. In the following, filtered quantities are denoted with a tilde, e.\,g.,
\begin{equation}
    \tilde{\rho}_\mathrm{err,v5X}^\mathrm{pre}(t) = \mathrm{HPF}(\rho_\mathrm{err,v5X}^\mathrm{pre}, \SI{0.08}{\milli\hertz})\ .
    \label{eq::rhoErrPrefit}
\end{equation}
The filtered pre-fit range errors for the three different \gls{LRI} data products (v51, v52, v53) are shown as blue traces in \cref{fig::postfit_rhoErr}. For saving computational costs, the range error is decimated to a sampling rate of \SI{3.3}{\milli\hertz}.
The signal in the pre-fit range error mainly oscillates at 1/rev and 2/rev frequencies, with varying amplitude over the months. The rms values for the traces are approximately \SI{25}{\micro\meter} for v51 and v52, and \SI{664}{\micro\meter} for v53. By estimating a global static scale $\epsSCF^\mathrm{glo}$ and time-shift $\zeta^\mathrm{glo}$, we can obtain post-fit residuals of the range error
\begin{equation}
	\tilde\rho^\mathrm{post}_\mathrm{err,v5X}(t) = \left( 1+\langle\epsSCF^\mathrm{glo}\rangle \right)\cdot\tilde\rho_\mathrm{LRI,v5X}^\mathrm{inst}(t+\zeta^\mathrm{glo}) - \tilde\rho_\mathrm{KBR}^\mathrm{inst}(t) - \tilde\rho_\mathrm{KBR}^\mathrm{FV}(t)\ ,
	\label{eq::rhoErrPostfit}
\end{equation}
which are significantly lower at the level of approx. \SI{6}{\micro\meter~rms} (cf. orange traces in \cref{fig::postfit_rhoErr}). 
\begin{table}[tb]
	\centering
	\caption{Global scale factor $\langle\epsSCF^\mathrm{glo}\rangle$ and timeshift $\zeta^\mathrm{glo}$ for the data span from 2018-Dec-13 to 2022-Jan-01. Here, prefit denotes the first step of the algorithm, before temperature sensors are added. Postfit denotes the parameters when all five temperature sensors were added.}
	\label{tab::TC_values_constants}
	\begin{tabular}{
			c
			S[table-format=3.3e1]
			>{\ }c<{\ }
			S[table-format=3.3e1]
			>{\ }c<{\ }}
		\hline
		LRI1B & {$\langle\epsSCF^\mathrm{glo}\rangle$} & $\zeta^\mathrm{glo}$ & {$\langle\epsSCF^\mathrm{glo}\rangle$} & $\zeta^\mathrm{glo}$ \\
		version & {postfit} & {postfit} & {postfit+TC} & {postfit+TC} \\
		\hline
		v51 & -1.992e-9 & \SI{71.15}{\micro\second} & -1.626e-8 & \SI{67.81}{\micro\second} \\
		v52 & -1.430e-9 & \SI{71.13}{\micro\second} & -3.810e-9 & \SI{67.95}{\micro\second} \\
		v53 &  2.235e-6 & \SI{71.11}{\micro\second} &  2.387e-6 & \SI{68.40}{\micro\second} \\
		\hline
	\end{tabular}
\end{table}
The estimated global parameters are given in columns 2 and 3 of \cref{tab::TC_values_constants} (without TC). They indicate that the high magnitude of the pre-fit range error was mainly caused by a static timeshift $\zeta^\mathrm{glo}\approx\SI{71}{\micro\second}$ between \gls{LRI} and \gls{KBR} in the case of v51 and v52, and by the scale (\SI{2.2}{ppm}) and timeshift in case of v53. These results were expected, e.\,g., the \SI{2.2}{ppm} scale offset was already apparent from \cref{fig::scales_SDS}.

We expect the \gls{KBR} noise level to limit the postfit range error. When assuming a $\SI{10}{\micro\meter\per\sqrt{\hertz}}$ white noise in the KBR at low Fourier frequencies and with a $\SI{3.3}{\milli\hertz}$ sampling rate, we obtain
\begin{equation}\label{eq::KBRnoiseLimit}
	\SI{10}{\micro\meter\per\sqrt{\hertz}}\cdot\sqrt{\SI{3.3}{\milli\hertz}/2} \approx \SI{0.4}{\micro\meter~rms}
\end{equation}	
as the \gls{KBR} noise limit, however the postfit range error is still above this level.

\begin{figure}[tb]
	\centering
	\subfloat[Scale factor $\langle\epsSCF\rangle$ for v51.]{
		\includegraphics[width=.475\linewidth]{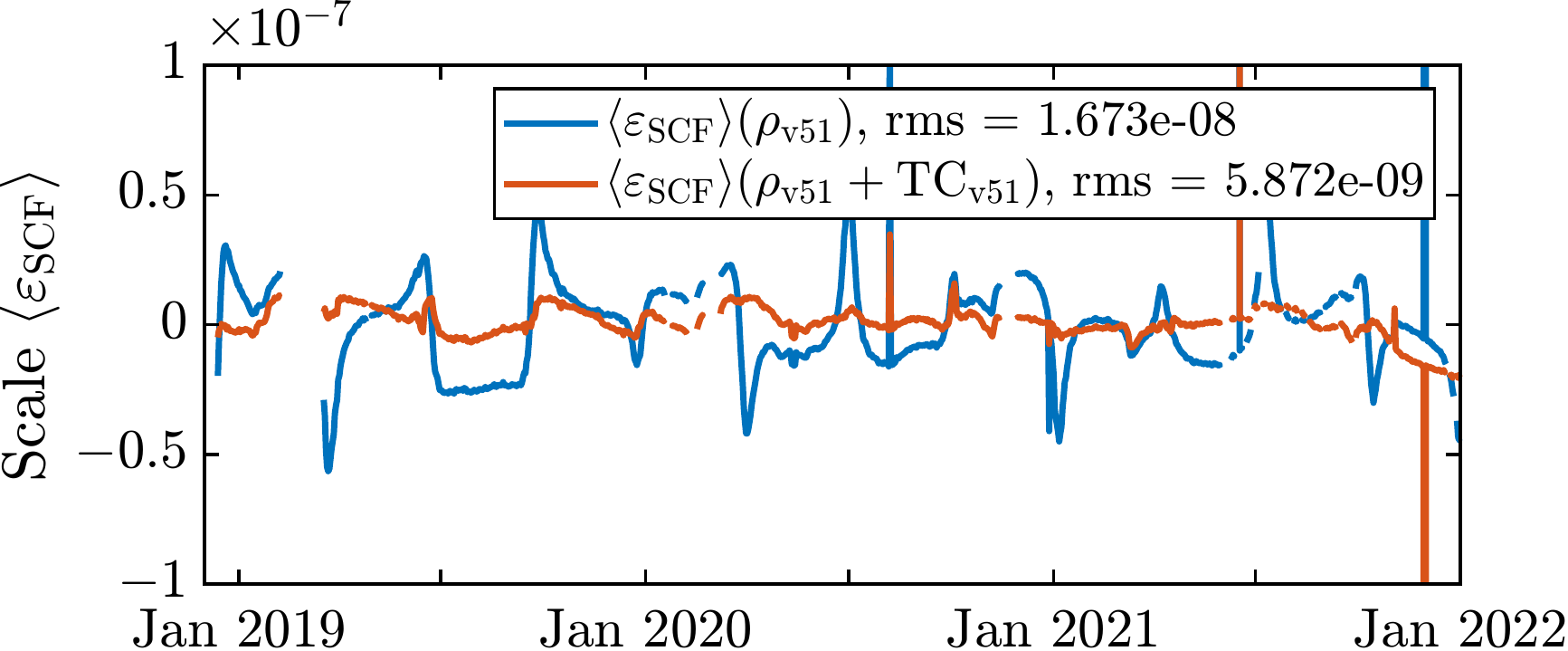}
	}
	\hfill
	\subfloat[Timeshift $\zeta$ for v51. Blue line shifted by \SI{75}{\micro\second}.]{
		\includegraphics[width=.475\linewidth]{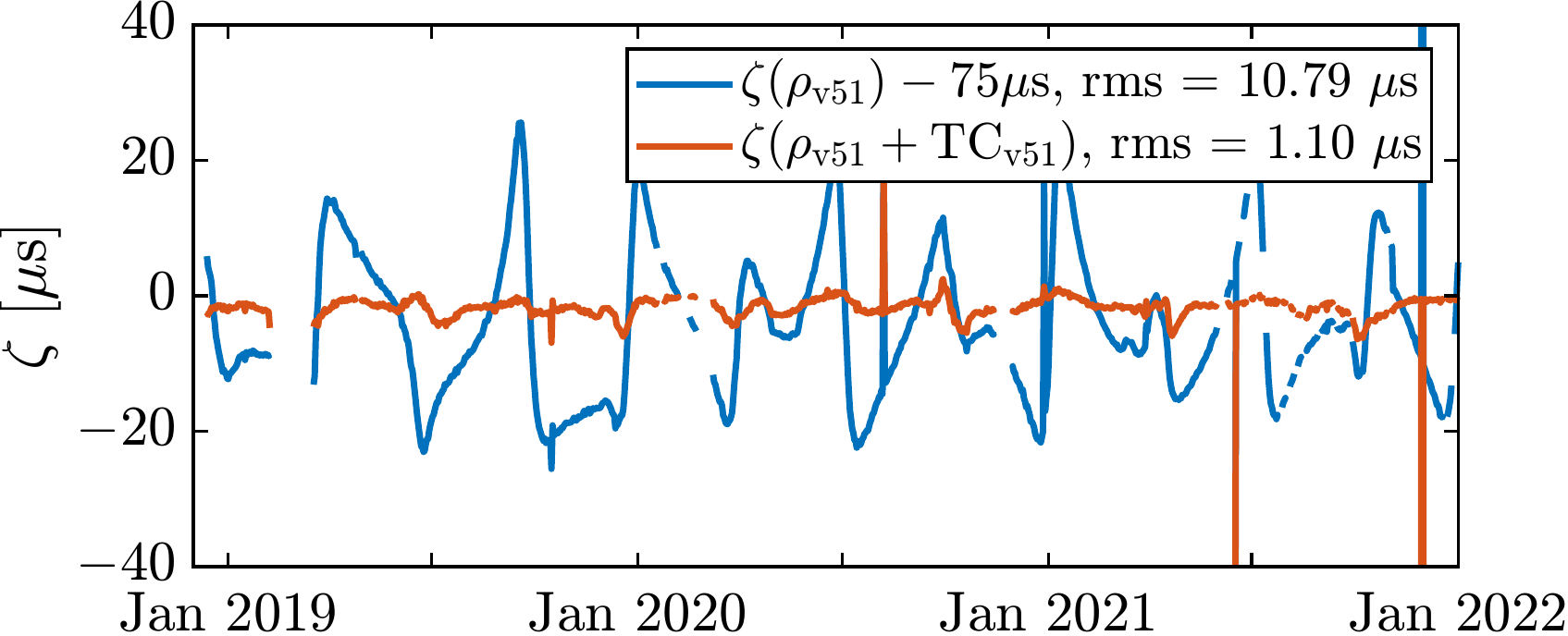}
	}

	\subfloat[Scale factor $\langle\epsSCF\rangle$ for v52.]{
		\includegraphics[width=.475\linewidth]{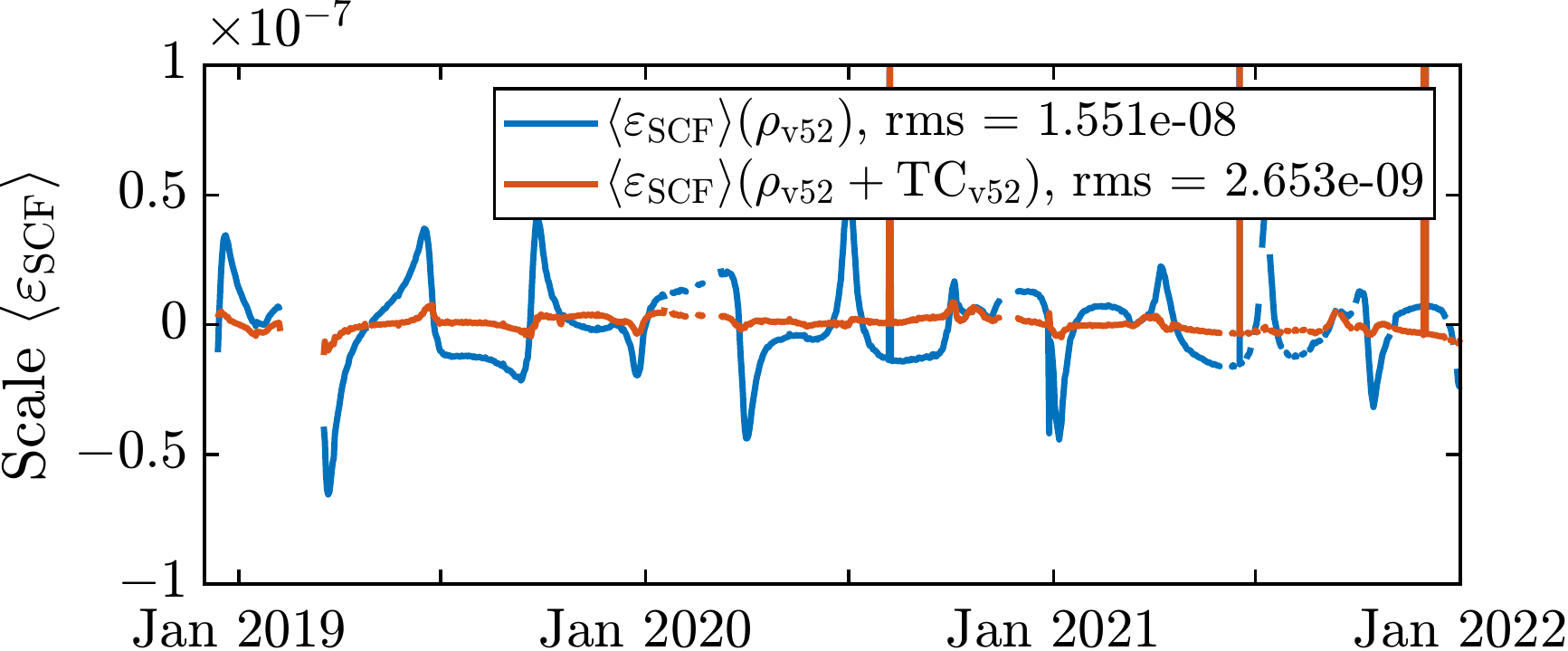}
	}
	\hfill
	\subfloat[Timeshift $\zeta$ for v52. Blue line shifted by \SI{75}{\micro\second}.]{
		\includegraphics[width=.475\linewidth]{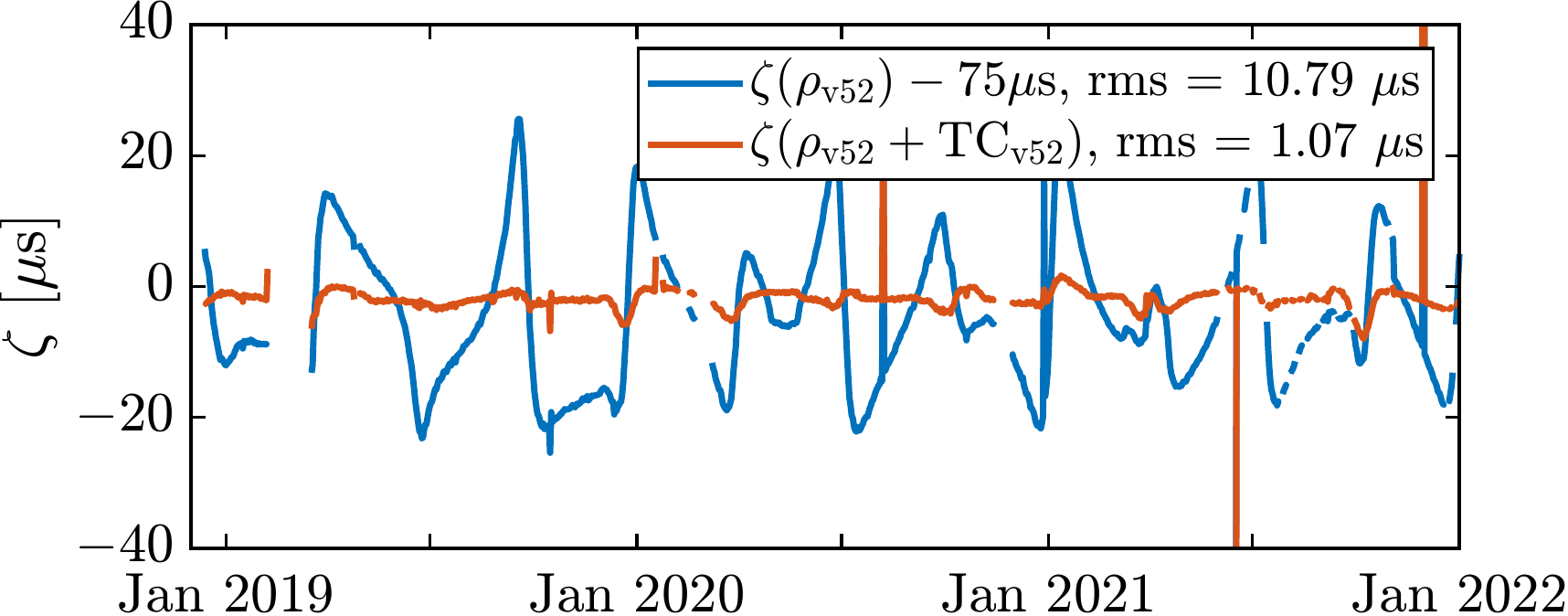}
	}

	\subfloat[Scale factor $\langle\epsSCF\rangle$ for v53. Blue line shifted by \num{-2.23e-6}.]{
		\includegraphics[width=.475\linewidth]{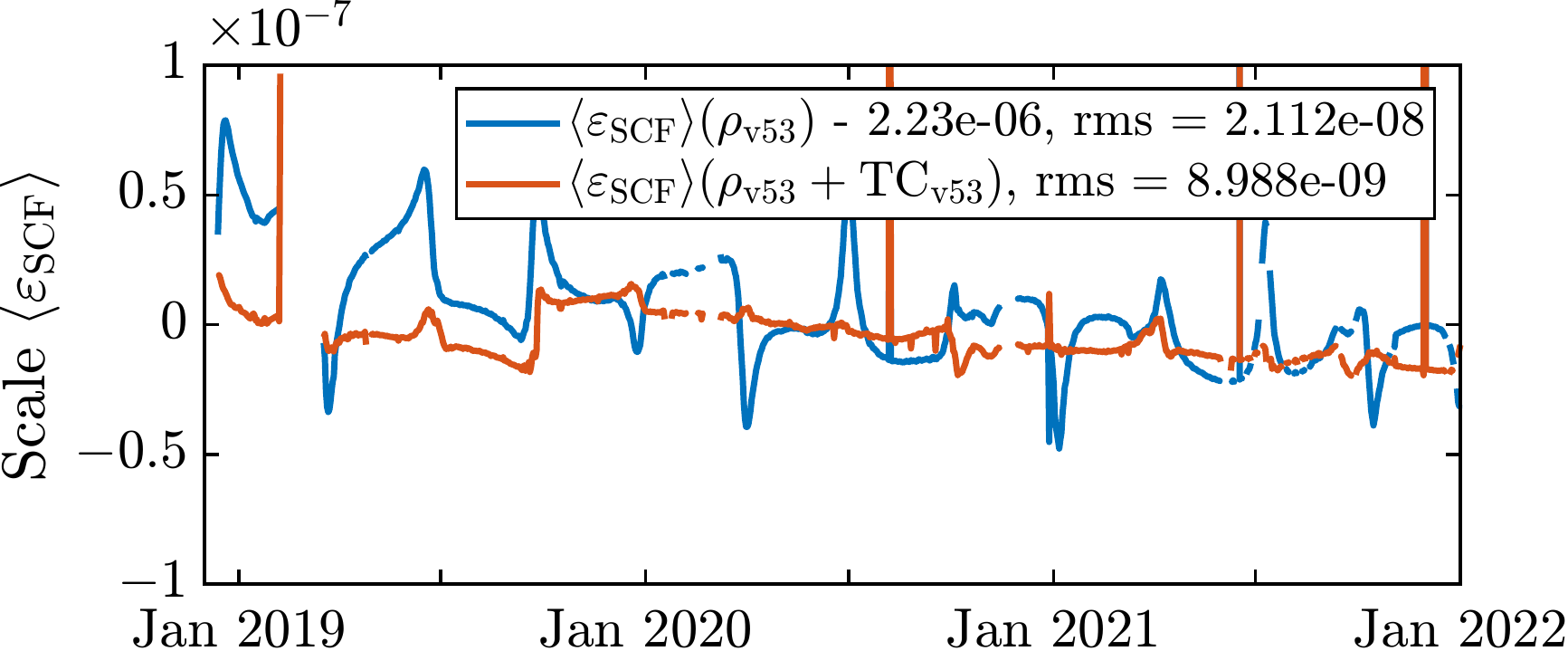}
		\label{fig::scales_nuAEIa}
	}
	\hfill
	\subfloat[Timeshift $\zeta$ for v53. Blue line shifted by \SI{75}{\micro\second}.]{
		\includegraphics[width=.475\linewidth]{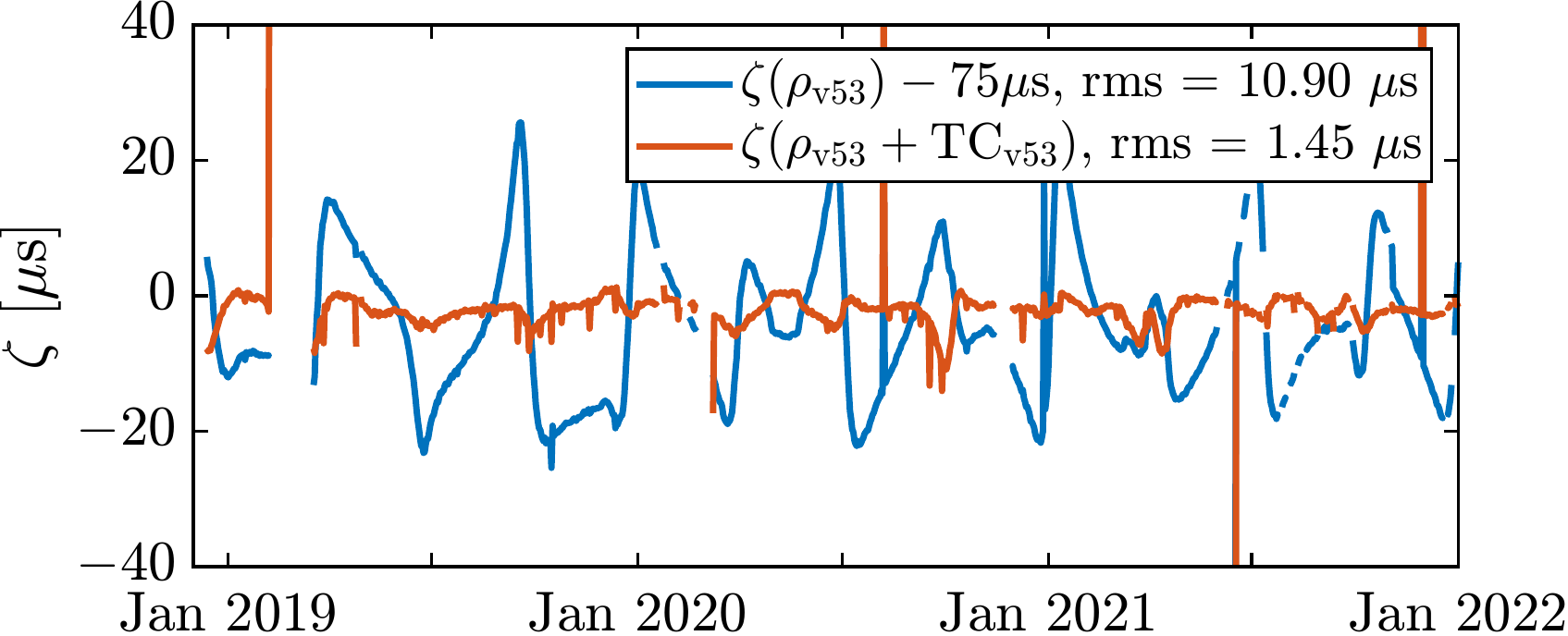}
		\label{fig::scales_nuAEIb}
	}
	\caption{Comparison of scale and timeshift for LRI1B-v51, v52 and v53. Each subplot shows the results from the raw data product (blue) and including the \glsfirst{TC} in red. Left column: Scale factor $\langle\epsSCF\rangle$ for different frequency models v51, v52, v53. Right column: corresponding timeshift $\zeta$. From top to bottom: v51, v52, v53. Outliers removed for computing the rms values.}
	\label{fig::scales_nuAEI}
\end{figure}

In addition to estimating and correcting a global mean scale and timeshift, we also estimated the scale and timeshift on a daily basis, which are shown as blue traces in \cref{fig::scales_nuAEI}.
Here, the scale factor shows some seasonal patterns with approx. 3 month period and with an amplitude of \SI{\pm0.05}{ppm} in all products; in v51 and v52 around zero, and in v53 around a \SI{2.2}{ppm} offset. The blue trace in the lower left panel for v53 also includes the exponential decay shown in \cref{fig::scales_SDS}. The timeshift in the right panels exhibits a \SI{75}{\micro\second} offset and seasonal variations with \SI{\pm25}{\micro\second} amplitude for all three products.

We note that the variations in the red traces of \cref{fig::postfit_rhoErr} with approx. \SI{6}{\micro\meter~rms} could be explained to a large extent with a daily varying scale and timeshift shown in \cref{fig::scales_nuAEI}. However, if these peaks and dips forming the seasonal variations with \SI{\pm0.05}{ppm} or equivalently \SI{\pm14}{\mega\hertz} amplitude are physical variations in the laser and the cavity resonance frequency, we would expect to see such variations in the laser telemetry and thus the telemetry-based laser frequency $\nu^\mathrm{TME}(t)$. We also lack an explanation for variations in the timeshift between \gls{KBR} and \gls{LRI}. A static timeshift could be produced by delays and uncertainties in the timing chain, though the exact contributor is not yet found (see \cref{app::timeref} for a brief discussion of the \gls{LRI} time frame). Therefore, in the next section, we investigate if the post-fit range error as defined by \cref{eq::rhoErrPostfit} can be further reduced when temperature coupling coefficients are co-estimated with the global scale and timeshift.

\section{\glsentrylong{TC} in KBR-LRI Residuals}\label{sec::toneErrors}
Changes in the thermal environment at many spacecraft units predominantly appear at 1/rev and 2/rev frequencies and may alter the measured range. We identified two possible coupling mechanisms. First, the coupling could be in the laser frequency regime, like temperature changes of the cavity or \gls{USO} acting as an additional scaling term. Second, errors could occur in the phase (pathlength) regime, e.\,g., due to temperature-dependent alignment of components or temperature-driven effects in the electronics.
In this section, we estimate linear coupling factors for different temperature sensors, with units of \si{1\per\kelvin} for the (fractional) frequency regime and \si{\meter\per\kelvin} for the phase regime, such that the residuals between \gls{LRI} and \gls{KBR} are further minimized. We call the sum of these two corrections the \glsreset{TC}\gls{TC}. The \gls{TC} coefficients and the global scale and time shift are estimated simultaneously so that the postfit residuals
\begin{equation}
	\tilde\rho_\mathrm{err,v5X}^\mathrm{post-TC}(t) = \left( 1+\langle\epsSCF^\mathrm{glo}\rangle \right)\cdot\tilde\rho_\mathrm{LRI,v5X}^\mathrm{inst}(t+\zeta^\mathrm{glo}) - \tilde\rho_\mathrm{KBR}^\mathrm{inst}(t) - \tilde\rho_\mathrm{KBR}^\mathrm{FV}(t) - \tilde\rho_\mathrm{TC}(t)
	\label{eq::rhoErrPostfitTC}
\end{equation}
are minimized.
We define the \gls{TC} as 
\begin{equation} \label{eq::TC}
	\tilde\rho_\mathrm{TC}(t) = \sum_i \delta\tilde\rho_{\mathrm{KBR-LRI},\,i}^\mathrm{freq}(t) + \sum_i \delta\tilde\rho_{\mathrm{KBR-LRI},\,i}^\mathrm{phase}(t) \ , 
\end{equation}
where we account for the two different error coupling mechanisms and $i$ denotes contributions from different temperature sensors $T_i(t)$.
In case of the frequency-domain coupling, we define
\begin{equation} \label{eq::TC_MHzK}
	\delta\tilde\rho_{\mathrm{KBR-LRI},\,i}^\mathrm{freq}(t) = 
	\mathrm{HPF}\left(L(t)\cdot \left( c_{1,i}\cdot T_i(t) + c_{2,i}\cdot\dot{T}_i(t) \right),\ \SI{0.08}{\milli\hertz}\right)\ .
\end{equation}
Importantly, we use a coupling where the scale error ($c_{1,i}\cdot T_i\approx\epsSCF$) is multiplied with the absolute distance $L\approx\SI{220}{\kilo\meter}$, cf. \cref{eq::delta-epsSCF}. 
The \si{\centi\meter}-precision of the \gls{GPS}-based absolute range $L$ obtained from GNI1B-v04 is sufficient here because the coupling coefficients are below \num{e-5}, which yields a precision of \SI{.1}{\micro\meter} or better. The coefficients $c_{1,i}$ and $c_{2,i}$ have units of \si{1\per\kelvin} and \si{\second\per\kelvin}, respectively. They can be converted to approximate equivalent laser frequency couplings in units of \si{\hertz\per\kelvin} by multiplying with $\nu_0\approx\SI{281}{\tera\hertz}$. The second term $c_{2,i}\cdot\dot{T}_i$ originates from a potential timeshift due to propagation time from temperature changes to the measurement and is linearized to first order. This timeshift can be computed by $\zeta_{T_i} = c_{2,i}/c_{1,i}$. It should be noted that a positive sign of $\zeta_{T,i}$ is not violating causality since the timeshift can always be regarded as a modulus w.\,r.\,t. the orbital frequency.

The phase-domain \gls{TC} contributors read
\begin{equation} \label{eq::TC_mK}
	\delta\tilde\rho_{\mathrm{KBR-LRI},\,i}^\mathrm{phase}(t) = 
	\mathrm{HPF}\left(c_{1,i}\cdot T_i(t) + c_{2,i}\cdot\dot{T}_i(t),\ \SI{0.08}{\milli\hertz}\right)\ ,
\end{equation}
where the coefficients $c_1$ and $c_2$ have the units \si{\meter\per\kelvin} and \si{\second\cdot\meter\per\kelvin}, respectively. The same high-pass filter used for the error range (cf. \cref{eq::rhoErrPrefit}) removes frequencies below 1 CPR, i.\,e., long-term drifts, but maintains 1 CPR, which has high relevance for the scale factor. 

We decompose the temperature $T_i$ into $T_i = T_{\mathrm{AC},i} + T_{\mathrm{DC},i}$ by high- and lowpass filtering, again using the same cutoff frequency of \SI{0.08}{\milli\hertz}. There are 161 thermistors $T_i$ per \gls{SC}, and the data is retrieved from so-called OFFRED data and downsampled to $\SI{3.3}{\milli\hertz}$ as well. We expect that the DC parts are more likely to cause variations in the frequency regime, while the AC parts cause \si{\meter\per\kelvin}-couplings. This is because the DC part contains a large static offset with only slight variations, which would imply a constant and hence irrelevant offset in $\tilde\rho_\mathrm{err}$, if the phase-domain coupling would apply, but prominent 1/rev tones in case of the frequency coupling due to the multiplication with $L$ in \cref{eq::TC_MHzK}.

An optimization algorithm iteratively picks a single temperature sensor that minimizes the postfit range error $\tilde\rho_\mathrm{err,v5X}^\mathrm{post-TC}$ the most. To do so, the parameters $c_{1,i}$ and $c_{2,i}$ from \cref{eq::TC_mK,eq::TC_MHzK} are determined for 
both components $T_{\mathrm{AC},i}$ and $T_{\mathrm{DC},i}$ of each temperature sensor $T_i$
in every iteration and the minimization gain, i.\,e., the residual rms of KBR-LRI \emph{with} this particular correction term divided by the residual rms \emph{without}, is computed.
For every iteration, the parameters of all the previously added sensors as well as $\langle\epsSCF^\mathrm{glo}\rangle$ and $\zeta^\mathrm{glo}$ are always co-estimated alongside the newly added sensor. Hence we extend the design matrix for the least squares minimization by two columns per iteration.

The algorithm stops after adding five sensors, giving 12 coefficients in total: two global scale $\langle\epsSCF^\mathrm{glo}\rangle$ and timeshift $\zeta^\mathrm{glo}$ biases for the whole period and two coefficients for each selected temperature sensor according to \cref{eq::TC,eq::TC_MHzK,eq::TC_mK}. The estimated constants for scale and timeshift are shown in the last two columns of \cref{tab::TC_values_constants}. The corresponding thermistor coefficients are shown in \cref{tab::TC_valuesa,tab::TC_valuesb,tab::TC_valuesc} for the ranging products v51, v52, and v53, respectively. 
The resulting residuals $\tilde\rho_\mathrm{err,v5X}^\mathrm{post-TC}$ are also shown in \cref{fig::postfit_rhoErr} (yellow). 
The subtraction of the full \gls{TC} model reduces the KBR-LRI rms residuals to a level of \SI{1.6}{\micro\meter} (v51), \SI{1.0}{\micro\meter} (v52) and \SI{2.3}{\micro\meter} (v53). Especially in the case of v52, this is close to the expected \gls{KBR} noise limit of \SI{0.4}{\micro\meter~rms} (cf. \cref{eq::KBRnoiseLimit}). 
\begin{table}[tb]
	\centering
	\caption{\glsentrylong{TC} parameters. The index $i$ denotes the order of importance, i.\,e., the gain in reducing the rms residuals. The $\delta\tilde\rho$ type denotes the coupling in phase or frequency regime. Thus, the unit of $c_1$ is \si{\meter\per\kelvin}, if the AC-component was used and \si{1\per\kelvin}, if the DC-component was used. The coefficient $c_2$ has units \si{\second\meter\per\kelvin} (AC) or \si{\second\per\kelvin} (DC). The last column $\zeta_T = c_2/c_1$ describes the timeshift of the temperature data in seconds. 
	}
	\label{tab::TC_values}
	\begin{adjustwidth}{-\extralength}{0cm}
		\raggedleft
		
	\subfloat[v51: Telemetry-based laser frequency model $\nu_{1/2}^\mathrm{TM}$]{
		\begin{tabularx}{0.99\linewidth}{*{2}{c}*{4}{Y}*{2}{R{0.13\linewidth}}R{0.08\linewidth}}
			\toprule
			$i$ & SC & Sensor ID & Ascii Name & $T_\mathrm{AC} / T_\mathrm{DC}$ & $\delta\tilde\rho$ type & \thead{$c_1$} & \thead{$c_2$} & \thead{$\zeta_T$ [\si{\second}]} \\
			\midrule
			1 & \GFC & THT10013 & SaMzPx & AC & phase & \num{-1.127e-07} & \num{-6.240e-06} & \num{55.4} \\
			2 & \GFD & THT10133 & BatTrp & AC & phase & \num{-1.173e-05} & \num{-1.379e-03} & \num{117.6} \\
			3 & \GFC & THT10144 & Pr21 & DC & freq & \num{-1.642e-09} & \num{-1.501e-06} & \num{914.0} \\
			4 & \GFC & THT10022 & LriLpcMy & DC & freq & \num{2.640e-09} & \num{1.182e-06} & \num{447.9} \\
			5 & \GFC & THT10143 & Oct11 & AC & phase & \num{4.153e-06} & \num{-9.814e-04} & \num{-236.3} \\
			\bottomrule
		\end{tabularx}
		\label{tab::TC_valuesb}}

	\subfloat[v52: Exponential cavity frequency decay model $\nu_\RR^\mathrm{Cav}$]{
		\begin{tabularx}{0.99\linewidth}{*{2}{c}*{4}{Y}*{2}{R{0.13\linewidth}}R{0.08\linewidth}}
			\toprule
			$i$ & SC & Sensor ID & Ascii Name & $T_\mathrm{AC} / T_\mathrm{DC}$ & $\delta\tilde\rho$ type & \thead{$c_1$} & \thead{$c_2$} & \thead{$\zeta_T$ [\si{\second}]} \\
			\midrule
			1 & \GFC & THT10013 & SaMzPx & AC & phase & \num{-1.302e-07} & \num{6.194e-06} & \num{-47.6} \\
			2 & \GFD & THT10138 & MepFrontPy & AC & phase & \num{-2.238e-07} & \num{5.403e-05} & \num{-241.4} \\
			3 & \GFC & THT10007 & GpsOccAnt & DC & freq & \num{1.513e-11} & \num{-2.382e-08} & \num{-1573.9} \\
			4 & \GFD & THT10089 & LriOba & AC & phase & \num{-1.080e-05} & \num{1.091e-03} & \num{-101.0} \\
			5 & \GFD & THT10113 & LriLas & DC & freq & \num{1.049e-10} & \num{-1.121e-06} & \num{-10689.7} \\
			\bottomrule
		\end{tabularx}
		\label{tab::TC_valuesc}}

	\subfloat[v53: Pre-calibrated fixed frequency value $\nu_0$]{
		\begin{tabularx}{0.99\linewidth}{*{2}{c}*{4}{Y}*{2}{R{0.13\linewidth}}R{0.08\linewidth}}
			\toprule
			$i$ & SC & Sensor ID & Ascii Name & $T_\mathrm{AC} / T_\mathrm{DC}$ & $\delta\tilde\rho$ type & \thead{$c_1$} & \thead{$c_2$} & \thead{$\zeta_T$ [\si{\second}]} \\
			\midrule
			1 & \GFD & THT10032 & SaMzMx & AC & phase & \num{-1.022e-07} & \num{1.923e-05} & \num{-188.2} \\
			2 & \GFD & THT10052 & AccPanel & DC & freq & \num{-9.162e-09} & \num{-4.827e-06} & \num{526.8} \\
			3 & \GFD & THT10138 & MepFrontPy & AC & phase & \num{-4.996e-06} & \num{8.787e-04} & \num{-175.9} \\
			4 & \GFD & THT10157 & Oct22 & DC & freq & \num{5.288e-10} & \num{3.349e-07} & \num{633.4} \\
			5 & \GFD & THT10052 & AccPanel & AC & phase & \num{8.433e-05} & \num{-3.014e-03} & \num{-35.7} \\
			\bottomrule
		\end{tabularx}
		\label{tab::TC_valuesa}}
	\end{adjustwidth}
\end{table}

\begin{figure}
	\centering
	\includegraphics[width=\linewidth]{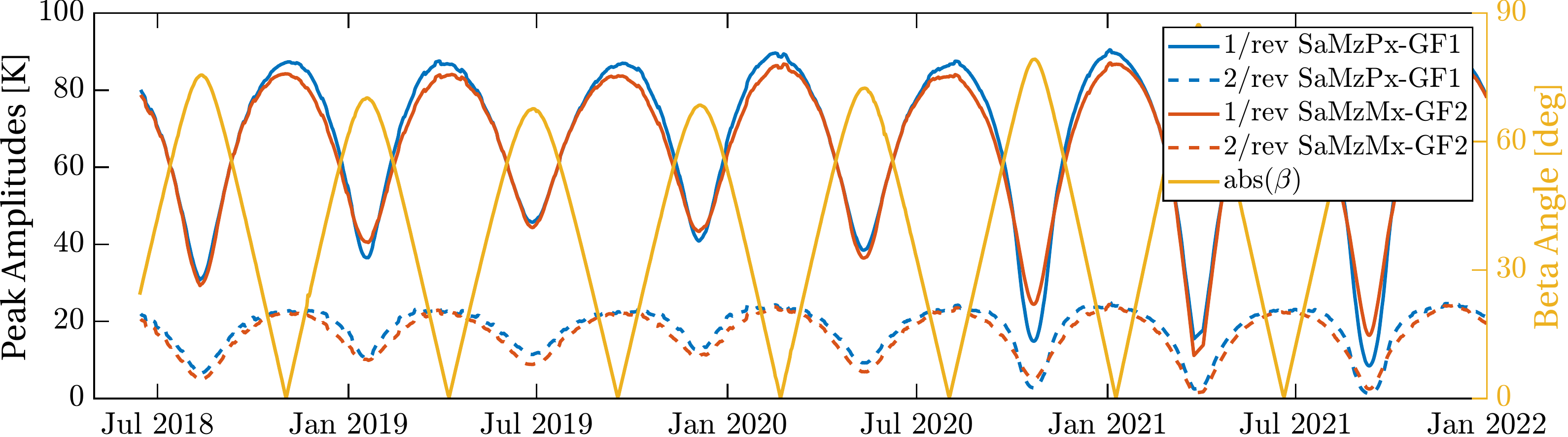}
	\caption{Peak amplitudes at 1/rev and 2/rev frequencies for SaMzPx of \GFC (blue) and SaMzMx of \GFD (orange), which are the most dominant \gls{TC} contributors. The yellow curve shows the absolute value of the $\beta$-angle between the orbital plane and the sun. The sinusoidal 1/rev and 2/rev temperature amplitudes are much higher when the $\beta$-angle is close to zero, i.\,e., when the sun is in the orbital plane.}
	\label{fig::SaMzPlot}
\end{figure}

We observe that in the first iteration, sensors called ``SaMz**'' were chosen in all three cases, which are attached to the zenith-pointing solar panels (\underline{s}olar \underline{a}rray \underline{m}inus $\underline{z}$). We expect that the underlying satellite interior's thermal environment is highly correlated to these sensors since the solar arrays are directly heated by the sun and thus exhibit large temperature variations. \Cref{fig::SaMzPlot} shows the dominant 1/rev and 2/rev amplitudes of these particular sensors.
A 1/rev amplitude of \SI{80}{\kelvin} results in a tone error of roughly \SI{8}{\micro\meter} at 1/rev, as apparent from the $c_1\approx \SI{0.1}{\micro\meter\per\kelvin}$ coupling factors in the first rows of \cref{tab::TC_valuesa,tab::TC_valuesb,tab::TC_valuesc}.

%
%
We highlight that most temperature sensors inside the spacecraft are highly correlated. Thus, there might be other sets of five sensors that could produce very similar results.

For verifying the \gls{TC}, the daily scale and timeshift of KBR and LRI are computed again using the LRI1B-v5X ranging products, but now we subtract the \gls{TC} model $\tilde\rho_\mathrm{TC}(t)$ beforehand. The new scale and timeshift are shown in orange also in \cref{fig::scales_nuAEI}. They clearly show less seasonal variations than the blue curves without \gls{TC}. 
The best performance, by means of reducing the variations in the daily scale factor estimate $\langle\epsSCF\rangle$ in the KBR-LRI differences, is achieved when using v52, which is based on the exponential cavity frequency model in combination with the \gls{TC}. Here, the rms variations of the scale factor $\langle\epsSCF\rangle(\rho_\mathrm{v52})$ are reduced from \num{1.551e-8} to \num{2.653e-9}. Also, the corresponding timeshift $\zeta(\rho_\mathrm{v52}+\mathrm{TC}_\mathrm{v52})$ exhibits low variations of about \SI{1}{\micro\second} when including the \gls{TC} correction $\tilde\rho_\mathrm{TC}$.

We emphasize that the \gls{TC} parameters shown in \cref{tab::TC_values_constants,tab::TC_valuesa} can also be used to correct the LRI1B-v04 dataset by SDS. However, one has to revert the effects of the daily $\langle\epsSCF^\mathrm{SDS}\rangle$ and $\zeta$ beforehand, which are already applied in LRI1B-v04. The timeshift $\zeta$ can be extracted by forming the difference of the time offsets (\texttt{eps\_time}) provided in CLK1B and LLK1B. 

If we assume that $\tilde\rho_\mathrm{TC}$ is mainly caused by the \gls{KBR} instrument, e.\,g., due to the stable cavity resonance frequency and thermally induced \gls{KBR} antenna phase center variations, the most precise inter-satellite ranging dataset is given by LRI1B-v52 with scale $\langle\epsSCF^\mathrm{glo}\rangle=\num{-3.810e-9}\approx\SI{-1}{\mega\hertz}$, timeshift $\zeta^\mathrm{glo}=\SI{67.95}{\micro\second}$ and without subtracting $\tilde\rho_\mathrm{TC}$.

Since the OFFRED thermistor data is not publicly available, we provide the \SI{0.5}{\hertz} ranging data product LRI1B-v52 and the corresponding \gls{TC} ranging correction called RTC1B-v52 (Range \glsentrylong{TC}), see Data Availability Statement below.

\section{Discussion of Results and Alternative Approaches for Future Missions}
\label{sec::discussion}
The frequency of the \gls{LRI} laser is needed to convert phase to range. Any error or uncertainty can be considered a scale error in the range measurement. Currently, the \gls{LRI} scale or absolute frequency is estimated daily by correlating \gls{KBR} range with \gls{LRI} range.
One goal of the methods presented in this paper is to derive an independent and reliable model for the absolute laser frequency $\nu_\RR$ of the \gls{LRI}.
Such models would be needed if \gls{KBR} data is missing, e.\,g., if the second \gls{IPU} on the \GFD satellite would become unavailable. Moreover, since the \gls{LRI} processing becomes less dependent on the \gls{KBR}, measurement errors in the \gls{KBR} would not affect the \gls{LRI} data anymore.
Furthermore, in a future mission, there will likely be only a single LRI-like ranging instrument which requires a new processing scheme. There are several options for determining the absolute laser frequency for these missions, which will be discussed briefly in the following.

The telemetry-based models $\nu^\mathrm{TME}_{1/2}$, which includes the empirical correction term $\nu^\mathrm{emp}_{1/2}$ from in-flight measurements, reached an accuracy of approximately $\SI{60}{\mega\hertz}\approx\SI{200}{ppb}$ (see \cref{fig::emp_fit_results}).
Since the laser is thermally coupled to the satellite platform, temperature variations of the surrounding units couple into the setpoint-based model but not into the true frequency determined by the cavity.
We emphasize that the \gls{LRI} is a technology demonstrator and the calibration of the laser frequency had only a moderate priority. However, the authors assume that such accuracy could also be achievable in a future mission from on-ground calibrations only if the laser is characterized more thoroughly, in particular, if the drift of \SI{40}{\mega\hertz\per\year} of the laser setpoint in $\nu^\mathrm{emp}_{1/2}$ is calibrated. Additionally, one might attempt to characterize the cavity frequency exponential decay on-ground and derive an estimate for the flight phase.

An alternative to determining the frequency from in-flight telemetry is to co-estimate it during gravity field recovery, as it is usually done for the accelerometer scales and biases \cite{Helleputte2009,Klinger2016}. However, any \gls{LRI} scale uncertainty mainly manifests at 1/rev and 2/rev frequencies, where the gravity field recovery processing strategies often introduce empirical parameters that partly absorb the scale factor \cite{Klinger2016}. Furthermore, the estimated scale is highly correlated with the $C_{2,0}$ spherical harmonics coefficient, which is mainly measurable at 1/rev and 2/rev frequencies and not estimated reliably in gravity field recovery \cite{Klinger2016}. We would expect that \gls{LRI} errors from a scale factor uncertainty would be at the level of the post-fit residuals of the gravity field recovery, which are much higher than the \gls{LRI} requirements of \num{e-7} or \num{e-8} (cf. \cref{sec::errorCouplingModel}).

A well-known and broadly used approach to obtain a well-defined absolute laser frequency relies on iodine spectroscopy, where the hyperfine transition line of an iodine molecule is used as an absolute reference for a laser lock \cite{Arie92}. This technology has also been used for calibrating the \gls{LRI} \glspl{RLU} (cf. \cref{app::ws6-iodine,app::ws7-iodine}), and there are activities ongoing to qualify such setups for the space environment, see e.\,g. \cite{Doeringshoff2017}. 
However, saturated doppler-free spectroscopy is likely incompatible with the available optical power from Tesat \glspl{RLU} used in the \gls{LRI} so far.
Hence, one would need to add optical amplifiers, which significantly increase the complexity and electric power consumption. Laboratory experiments showed absolute frequency repeatability levels below $\SI{1}{\kilo\hertz}\approx\SI{3e-3}{ppb}$ \cite{Doeringshoff2017}. Thus, this method is probably the most accurate mean to fix or determine the scale factor.
Optionally, there is the possibility for a hybrid lock by using both a conventional \gls{PDH} lock to an optical cavity and a spectroscopic locking to an iodine reference. This hybrid lock combines the stability of the cavity at high Fourier frequencies, and the absolute laser frequency knowledge through the molecular reference \cite{Sanjuan2021}. 

Another approach for measuring the cavity resonance frequency is based on an extension of the \gls{PDH} lock \cite{DeVoe1984}. Adding an additional tone (scale factor tone) at a few \si{\mega\hertz} frequency together with upper and lower sidebands with the approximate \gls{FSR} separation enables the readout of the actual cavity \gls{FSR} w.\,r.\,t. the applied sidebands. The frequency of these sidebands is derived coherently from the \gls{USO} as the two tones and \gls{LRP} time. After determination of the \gls{USO} frequency during precise orbit determination, this technique can provide an estimate for the \gls{FSR} frequency of the cavity. The cavity resonance frequency and \gls{FSR} are linearly related to each other through $\nu = n_\mathrm{mode}\cdot\mathrm{FSR}+\mathrm{offset}$, where the mode number $n_\mathrm{mode}$ is an integer mode number. The offset must be calibrated on ground \cite{Rees2022}.
The principle has been demonstrated in laboratory experiments with an accuracy of roughly $\SI{3}{\mega\hertz}\approx\SI{10}{ppb}$. It is noteworthy that the scale factor tone and its sidebands have little influence on the conventional \gls{PDH} readout \cite{Rees2021,Rees2022}. The advantage of this technique is that only minor changes to existing flight hardware are needed, e.\,g., the use of \si{\giga\hertz} electro-optical modulators instead of \si{\mega\hertz}. However, additional RF electronics and an additional processing unit for the readout are required. The \gls{FSR}-readout is currently the most probable solution for upcoming gravity missions.

\section{Conclusion}
\label{sec::conclusion}
In this paper, the methodology as presented by \cite{Mueller2022} to derive a precise range from raw interferometric phase measurements was applied to in-flight data of the \gls{GFO} \gls{LRI} instrument. 
Based on that work, we derived the two dominant error terms, namely a time-variable scaling of the laser frequency, expressed through a scale factor $\epsSCF$, and a timeshift $\zeta$ of the \gls{LRI} measurement w.\,r.\,t. the reference measurement is given by the \gls{KBR}. Importantly, variations in $\epsSCF$ couple into the range proportionally to the absolute distance $L$ between the two satellites. The scale $\epsSCF$ and timeshift $\zeta$ parameters can easily be compromised if errors in the range measurement at 1/rev and 2/rev frequencies are present.

In the second part, three different models to calculate the in-flight laser frequency were shown, which are largely independent of \gls{KBR} measurements, once the model parameters are determined. Based on these models, we derive three versions of an LRI1B-equivalent data product, namely v51, v52, and v53. The first method (v53) uses a constant, nominal laser frequency $\nu_0$, which was chosen early before launch, and without sophisticated analysis since it was clear that the numerical values serve as a start value for the subsequent and accurate refinement utilizing cross-calibration. Thus, v53 is, in principle, a pre-release for the official LRI1B-v04 dataset. The latter is further refined by daily cross-calibrating \gls{LRI} and \gls{KBR} ranging data, i.\,e.\,, estimating scale $\langle\epsSCF\rangle$ and time-shift $\zeta$ on a daily basis. The daily scale factor or frequency $\nu^\mathrm{SDS}$ determined from the cross-calibration revealed seasonal variations and a settling effect, which we attribute to the optical reference cavity and which might be related to aging of the \gls{ULE} cavity spacer material as reported in \cite{Alnis2008}.
We use an exponential decay function (cf. \cref{eq::epsCav}) to describe this settling effect and to form v52. The v51 dataset uses a laser frequency model derived from \gls{LRI} laser and temperature telemetry. 
One can relate the lasers' control loop setpoints and temperatures to the output frequency using linear coupling factors, which were calibrated on ground before launch. The setup of the pre-flight calibration measurements was explained, and the calibration factors were provided. The initial $\nu^\mathrm{TM}$-model from on-ground calibrations was then compared to four years of in-flight data of the \gls{GFO} mission. It was found that the $\nu^\mathrm{TM}$ model frequency or, more precisely, the setpoints of the laser control loops drift over time by roughly \SI{40}{\mega\hertz\per\year}. Furthermore, two steps were observed when the lasers were operated in non-nominal conditions. The physical reason for the drift could be a consequence of aging effects of the \gls{NPRO} crystal or electronics. However, while the exact reason remains unknown, we compensate for the drift and steps with an empirical model here. We observed that this telemetry-based model $\nu^\mathrm{TME}_{1/2}=\nu^\mathrm{TM}_{1/2}-\nu^\mathrm{emp}_{1/2}$ of the laser frequency does not show seasonal variations, indicating that the seasonal variations of the frequency $\nu^\mathrm{SDS}$ are not actual changes in the laser frequency. The v51 dataset uses this empirically corrected model $\nu^\mathrm{TME}_{1/2}$.

Afterward, we focused on analyzing residuals of the difference of KBR-LRI, which we call the range error. At first, the direct difference yields large prefit range errors $\tilde\rho_\mathrm{err,v5X}^\mathrm{pre}$ of more than \SI{25}{\micro\meter~rms} for all three v5X data sets. In the second step, the effect of a global scale factor $\langle\epsSCF^\mathrm{glo}\rangle$ and a global timeshift $\zeta^\mathrm{glo}$ are subtracted, reducing the postfit residuals $\tilde\rho_\mathrm{err,v5X}^\mathrm{post}$ to approximately $\SI{6}{\micro\meter~rms}$ in all three cases. 

These postfit residuals could be explained by seasonal variations in the scale and timeshift as determined from daily cross-calibration of \gls{LRI} w.r.t. \gls{KBR}. 
However, since the telemetry-based frequency model $\nu^\mathrm{TME}$ does not show these seasonal variations, we described them with a \glsfirst{TC}. We accounted for two \gls{TC} mechanisms, one in the phase domain and one in the frequency domain. An algorithm to determine coupling coefficients for all temperature sensors on both spacecraft was explained, and equations to compute the \gls{TC} were given. 
For each of the three datasets, a \gls{TC} composed of 12 coefficients, which includes five temperature sensors, was derived. For each temperature sensor, we derived a linear coupling $c_1$ with units of \si{\meter\per\kelvin} (phase domain) or \si{1\per\kelvin} (frequency domain) and a possible time delay $\zeta_T$. Furthermore, the two global parameters for the scale $\langle\epsSCF^\mathrm{glo}\rangle$ and timeshift $\zeta^\mathrm{glo}$, which are constant over the whole mission span, were refined. We showed that the differences between \gls{LRI} and \gls{KBR} can be reduced from approx. \SI{25}{\micro\meter~rms} to \SI{1}{\micro\meter~rms} when using LRI1B-v52 including the \gls{TC}. 
In all three cases, the dominant thermal coupling originates from thermistors attached to the zenith-facing solar array. An analysis of these thermistor timeseries' revealed that their 1/rev and 2/rev amplitudes are highest when the angle $\beta$ between the orbital plane and the sun is close to zero, which occurs roughly every six months. The tone error magnitude of these sensors in the \gls{TC} is in the order of \SI{\pm 8}{\micro\meter}.

We analyzed only the thermal effects that are not common in \gls{LRI} and \gls{KBR}, i.\,e., which appear in the KBR-LRI difference. Thus, the \gls{TC} does not address potential common effects.

This paper introduced a new laser frequency model for the \gls{LRI}, representing the current best knowledge of the cavity resonance frequency on \GFC. We showed that this resonance frequency is relatively stable after an initial exponential convergence. As apparent from daily KBR-LRI cross-calibration, seasonal variations can be explained with tone errors from a \glsentrylong{TC}.

\vspace{6pt} 

\authorcontributions{Conceptualization: Malte Misfeldt and Vitali Müller; Funding acquisition: Gerhard Heinzel; Investigation: Malte Misfeldt, Vitali Müller, Laura Müller and Henry Wegener; Project administration: Gerhard Heinzel; Writing -- original draft: Malte Misfeldt; Writing -- review \& editing: Vitali Müller, Laura Müller, Henry Wegener and Gerhard Heinzel.}

\funding{This work has been supported by: The Deutsche Forschungsgemeinschaft (DFG, German Research Foundation, Project-ID 434617780, SFB 1464); Clusters of Excellence ``QuantumFrontiers: Light and Matter at the Quantum Frontier: Foundations and Applications in Metrology'' (EXC-2123, project number: 390837967); the European Space Agency in the framework of Next Generation Geodesy Mission development and ESA's third-party mission support for GRACE-FO; the Chinese Academy of Sciences (CAS) and the Max Planck Society (MPG) in the framework of the LEGACY cooperation on low-frequency gravitational-wave astronomy (M.IF.A.QOP18098).}

\dataavailability{GRACE-Follow On Level-1 instrument data is distributed by the GRACE/GRACE-FO project via NASA PODAAC (\url{https://podaac-tools.jpl.nasa.gov/drive/files/allData/gracefo}). 
The data product LRI1B-v52 as derived in this manuscript, together with the Range Thermal Coupling product RTC1B-v52 is available at the LUH data repository \cite{data_LRI1Bv52}.
Thermistor data from OFFRED telemetry are not publicly available by the time of writing.}

\acknowledgments{The authors would like to thank the JPL LRI team for helpful regular discussions and insights.}

\conflictsofinterest{The authors declare no conflict of interest.} 

\abbreviations{Abbreviations}{
	The following abbreviations are used in this manuscript:
	\renewcommand{\glossarysection}[2][]{}
	\printglossary[type=\acronymtype,title=Abbreviations,nopostdot,nonumberlist,nogroupskip]
}

\appendixtitles{yes} 
\appendixstart
\appendix
\glsresetall

\section[\appendixname~\thesection]{Calibration of WS6-600 using an Iodine Cell}
\label{app::ws6-iodine}
Various measurement campaigns for determining the \gls{TM} frequency models for the two laser flight models of the \gls{LRI} have been performed between July 2017 and January 2018. In these campaigns, three different wavelength meters (or wavemeters) have been used: A WS6-600 and a WS7-60 by HighFinesse / Angstrom and a WA1500 by Burleigh. The first one has an absolute accuracy of \SI{600}{\mega\hertz}, while the latter two are more accurate by one order of magnitude. 
The two HighFinesse devices were used mainly during the testing campaigns. They have a built-in calibration via a neon lamp. The reference frequency of a well-known iodine hyperfine transition was used to verify this internal calibration. Iodine is a commonly used molecule for stabilizing lasers to an absolute frequency reference, see e.\,g. \cite{Arie92,Schuldt2003}.

At first, the accuracy of the WS6-600 was measured in a setup utilizing a Prometheus laser by Coherent, Inc. \cite{Prometheus_DataSheet} providing \SI{500}{\milli\watt} output power at a wavelength of \SI{1064}{\nano\meter}. The secondary output provides \SI{20}{\milli\watt} of frequency-doubled light at \SI{532}{\nano\meter}. 
The frequency of this reference laser can be tuned over a range of roughly \SI{60}{\giga\hertz} via thermal elements, and piezoelectric transducers \cite{Prometheus_DataSheet}.

The laser's frequency was locked via Doppler-free \gls{MTS} to the \mbox{R(56)32-0} iodine line, of which we used the $a_1$ and $a_{10}$ components. The iodine cell was manufactured by InnoLight as well. The nominal frequency of the $a_{10}$ hyperfine component is \cite{Riehle2018}
\begin{equation}
\nu_{a_{10}} = \SI{563260223.513}{\mega\hertz}\ .
\end{equation}
The $a_{10}$ component's frequency is elevated by $\delta\nu_\text{vis} = \SI{572.1}{\mega\hertz}$ w.\,r.\,t. the $a_1$ component \cite{Arie92}. Hence,
\begin{equation}
\nu_{a_{1}} = \nu_{a_{10}}-\delta\nu_\text{vis} = \SI{563259651,413}{\mega\hertz}\ .
\end{equation}
The corresponding difference frequency at \SI{1064}{\nano\meter} is
\begin{equation}
\delta\nu_\text{IR} = \SI{286.05}{\mega\hertz}\ ,
\end{equation}
which would be the accurate result.
\begin{figure}[tb]
	\begin{minipage}[t]{0.48\linewidth}
		\includegraphics[width=\linewidth]{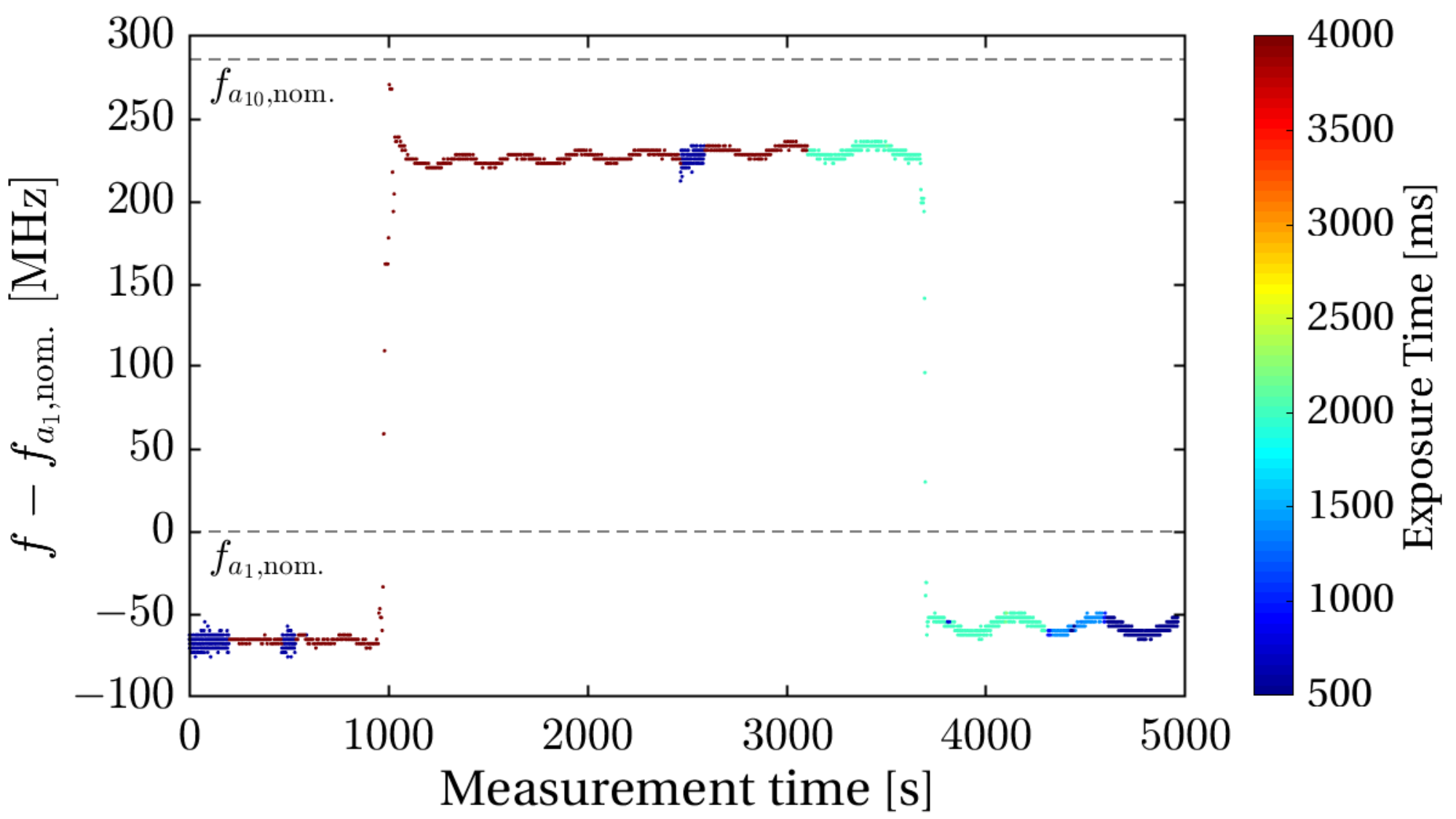}
		\caption{Absolute frequency measurements of the reference laser locked to different hyperfine lines of an iodine cell. The apparent quantization of approximately \SI{3}{\mega\hertz} arises from the finite resolution of \SI{e-5}{\nano\meter} of the wavemeter WS6-600. Outliers removed. The small oscillations are observed repeatedly for this wavemeter. The color indicates the wavemeter's exposure time.}
		\label{fig::WS6-600_longterm}
	\end{minipage}
	\hfil
	\begin{minipage}[t]{0.48\linewidth}
		\includegraphics[width=\linewidth]{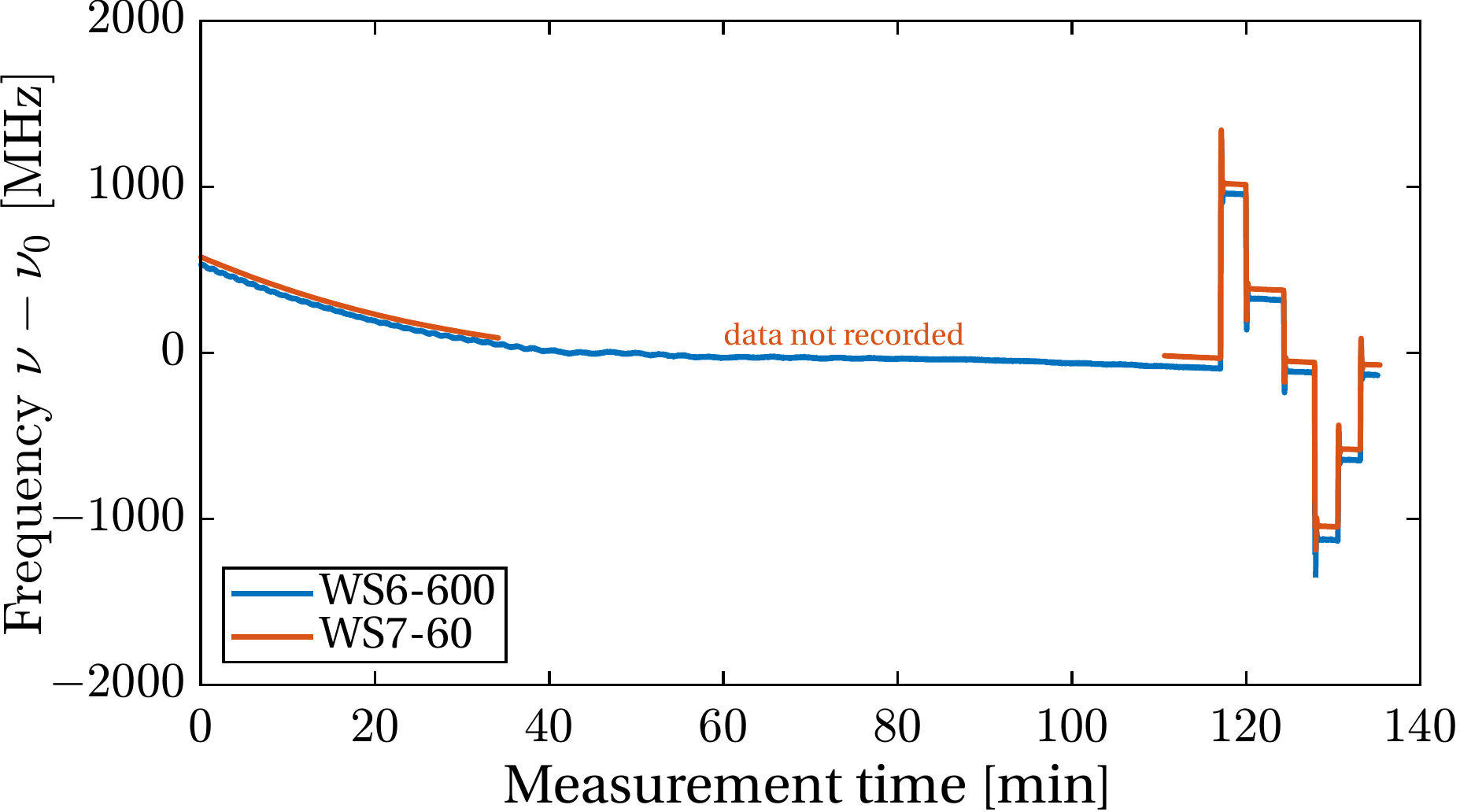}
		\caption{OGSE laser frequency, measured with two wavemeters. \SI{110}{\minute} waiting for thermal equilibrium, afterwards active control using the laser's thermal elements. Outliers removed. $\nu_0 = \SI{281616307}{\mega\hertz}$.}
		\label{fig::GSE_laser_calib}
		\vfill
	\end{minipage}
\end{figure}
Our first calibration measurement, shown in \cref{fig::WS6-600_longterm}, took \SI{5000}{\second}, of which the laser was locked to the $a_1$ component in the beginning and the end for approximately \SI{1000}{\second} and the $a_{10}$ component in between. The optical power in the fiber going to the WS6-600 was relatively low at about \SI{850}{\nano\watt}. The average frequency measured for $\nu_{a_1}$ at \SI{1064}{\nano\meter} is about \SI{60}{\mega\hertz} below the nominal value, which is within the \SI{600}{\mega\hertz} accuracy of the device. The measured frequency difference between the two hyperfine components $a_1$ and $a_{10}$ is $\delta\nu_\mathrm{meas.}=\SI{288.785}{\mega\hertz}$, being \SI{2.735}{\mega\hertz} higher than $\delta\nu_\mathrm{IR}$. Other measurements confirmed a bias of this device by approximately 20 to \SI{60}{\mega\hertz}, while the relative measurements are more precise.
The small oscillations with a magnitude of up to \SI{15}{\mega\hertz} and a period of about \SI{300}{\second} have been observed repeatedly for the WS6-600. The measurement is often noisier at short exposure times of about \SI{600}{\milli\second} (cf. color encoding in \cref{fig::WS6-600_longterm}).

\section[\appendixname~\thesection]{Calibration of OGSE laser and WS7-60}
\label{app::ws7-iodine}
After calibrating the WS6-600 wavemeter, the laser of the \gls{OGSE} was characterized w.\,r.\,t. its thermal coupling and drifts. The \gls{OGSE} laser was used in single-spacecraft functional tests to simulate the received light for the \gls{LRI} units. The flight laser units were operated in transponder mode and were locked to the incoming \gls{OGSE} light. At the integration facility, a second wavemeter, the WS7-60 with an absolute accuracy of \SI{60}{\mega\hertz}, was available alongside the WS6-600. Hence, this allowed us to calibrate the WS7-60 against the WS6-600.

A comparison of the two wavemeters, shown in \cref{fig::GSE_laser_calib}, revealed that the WS6-600 measures frequencies which are lowered by approximately \num{40} to \SI{80}{\mega\hertz} compared to the WS7-60. This is consistent with the calibration with iodine lines (see \cref{app::ws6-iodine,fig::WS6-600_longterm}). It was found that the \gls{OGSE} laser needs at least 60 to \SI{80}{\minute} to reach thermal equilibrium. The frequency drift after this warm-up phase is below \SI{20}{\kilo\hertz\per\second}. Since the two measurements agree well and the WS6-600 offset was observed before, it was concluded that the WS7-60 is accurate within the needs.

\section[\appendixname~\thesection]{Thermal actuator signals of \GFC}
\label{app::thermIlOol}
\begin{figure}[ht]
	\subfloat[\GFC actuator signals. ]{ %
		\centering
		\includegraphics[width=0.48\linewidth]{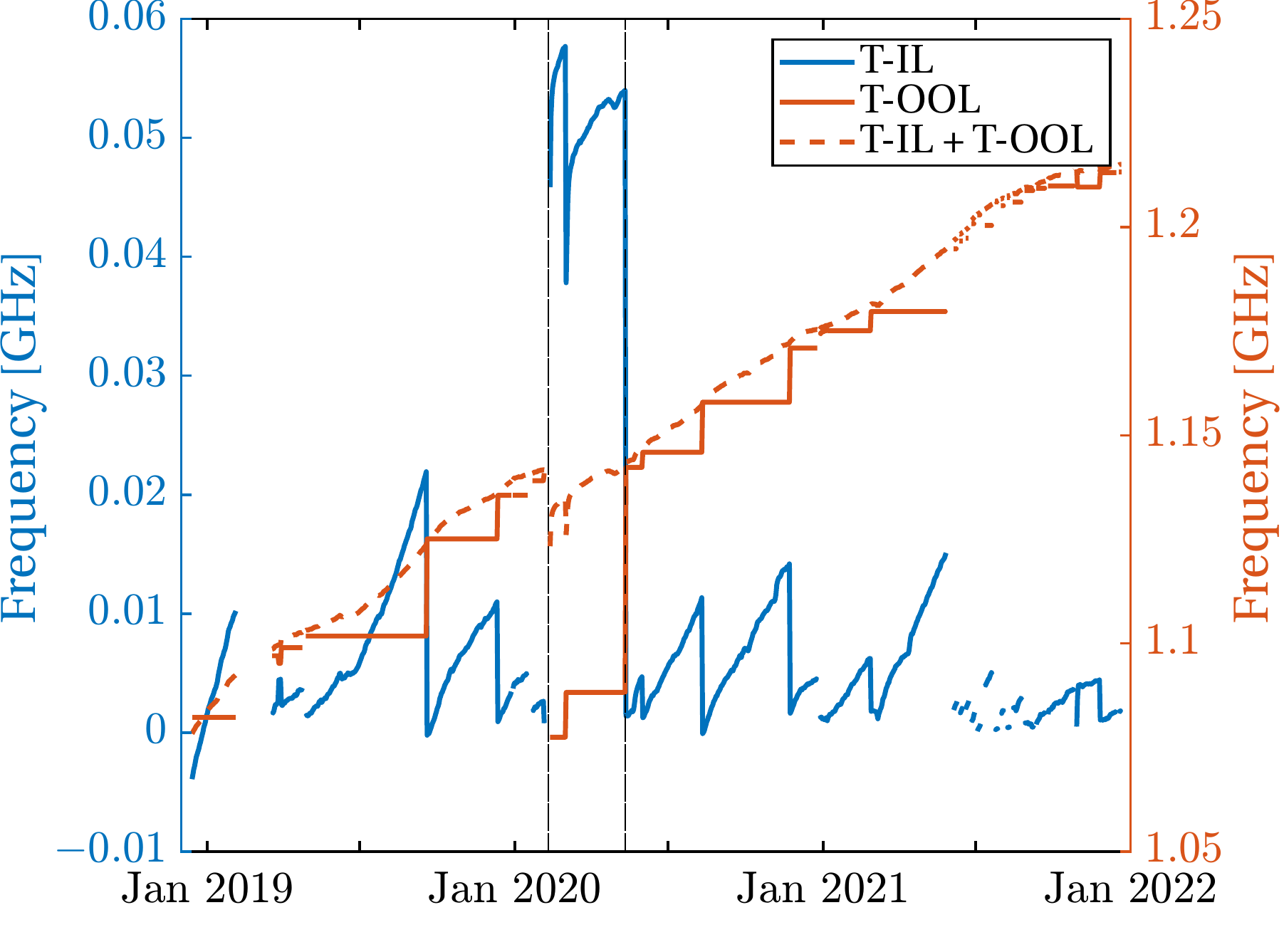}
	}
	\hspace{0.01\linewidth}
	\subfloat[\GFD actuator signals. ]{
		\centering
		\includegraphics[width=0.48\linewidth]{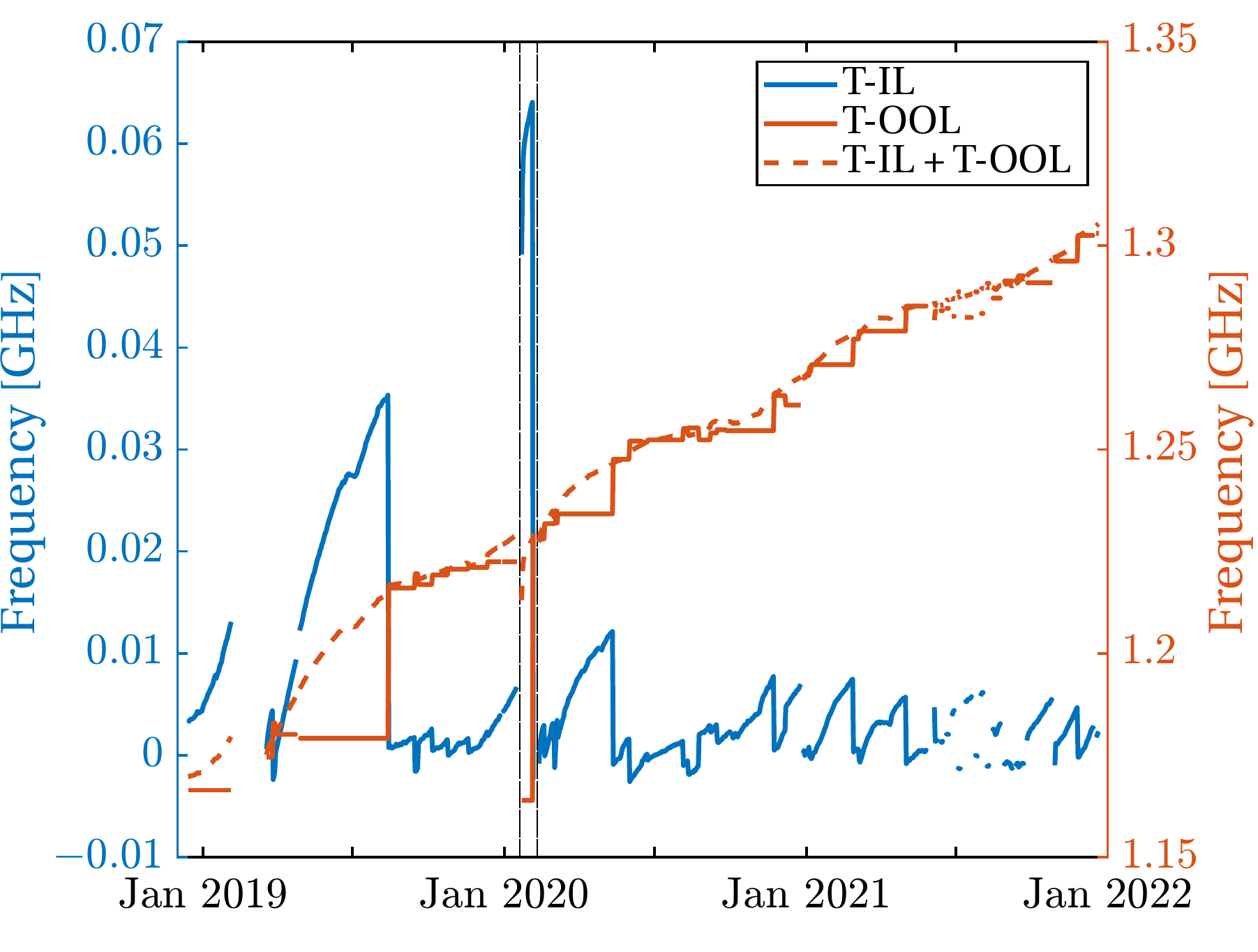}
	}
	\caption{Control loop actuator signals multiplied with their calibrated coefficient (see \cref{tab::designvalues}) for the thermal actuator of the \glspl{RLU} from 2018-Dec until 2022-Jan. \emph{Left axes:} \Gls{IL} signal (blue). \emph{Right axes:} \Gls{OOL} signal (orange) and sum of both (dashed orange). The black vertical lines indicate the times, where a step was applied in the empirical model. \emph{(a)} A step is visible at the first dashed vertical line, where the LRI locked before reaching thermal equilibrium. At the second vertical line, no frequency step is visible, but the actuator signals jumped back to the usual range. \emph{(b)} Similar but smaller steps as in the left subplot are present between the two dashed vertical lines.}
	\label{fig::thermIlOol}
\end{figure}

\Cref{fig::thermIlOol} illustrates the in-flight thermal actuator signals of the two LRI units. 
The black vertical lines highlight the regions, where the \gls{LRI} acquired the link before the thermal equilibrium of the lasers was reached. This caused the in-loop signal to reach higher values than usual. At these instances, the telemetry-based laser frequency model shows nonphysical steps, which are then corrected by empirical parameters in \cref{eq::driftModel}.

\section[\appendixname~\thesection]{LRI Time Frames}\label{app::timeref}
The \gls{LRI} time frame is initialized at the startup of the \gls{LRP}, however, this initialization introduces an unknown offset of \SI{1.5}{\second} at maximum between the \gls{OBC} time and the \gls{LRI} time. After initialization, the \gls{LRP} clock is counting eight-fold \gls{USO} ticks, which implies a clock rate of \SI{38.656000}{\mega\hertz} for \GFC and \SI{38.656792}{\mega\hertz} for \GFD.
The offset between the time frames of the \gls{LRI} and the \gls{OBC} is regularly determined using so-called datation reports. The reported datation bias usually remains constant between reboots of either the \gls{LRP} or the \gls{IPU}. Furthermore, a filter is used inside the \gls{LRP} to reduce the phase telemetry sampling rate to roughly \SI{10}{\hertz}. This filter introduces a delay of \SI{28802038}{clock ticks} (approximately \SI{0.75}{\second}, slightly different for the two spacecraft) that has to be accounted for \cite{Ware_2018_DatationTest,Level1UserHandbook}. In actual flight-data processing, when \gls{KBR} and \gls{LRI} range data are cross-calibrated, an estimated additional offset of $\zeta\approx\SI{70}{\micro\second}$ is observed, whose origin is unknown and that is not measured with \gls{LRI} datation reports. During the analysis presented in this paper, this time shift $\zeta$ is numerically estimated in every comparison between \gls{LRI} and \gls{KBR} ranging data.


\begin{adjustwidth}{-\extralength}{0cm}

\reftitle{References}
\bibliography{scf_determination}
\end{adjustwidth}
\end{document}